\documentclass[twocolumn]{aastex631}

\usepackage{amsmath}
\usepackage{mathtools}
\usepackage{comment}

\submitjournal{ApJ}

\shorttitle{MELTYQ}

\begin{document}

\title{Coupling magma-ocean and atmospheres in spectral retrievals of sub-Neptunes
}

\footnote{\today}

\correspondingauthor{Yuichi Ito}
\email{yuichi.ito.kkyr@gmail.com}

\author[0000-0002-0598-3021]{Yuichi Ito}
\affiliation{Astrobiology Center, National Institutes of Natural Sciences (NINS), 2-21-1 Osawa, Mitaka, Tokyo 181-8588, Japan} 
\affiliation{Division of Science, National Astronomical Observatory of Japan, 
NINS, 2-21-1 Osawa, Mitaka, Tokyo 181-8588, Japan}

\author[0000-0001-6516-4493]{Quentin Changeat}
\affiliation{Kapteyn Institute, University of Groningen, 9747 AD Groningen, The Netherlands}

\begin{abstract}
Recent high-precision atmospheric observations with JWST is enabling detailed characterization of sub-Neptune atmospheres and motivating efforts to understand and constrain 
their interiors. Theoretical studies suggest that sub-Neptunes possibly host
hydrogen-dominated atmospheres that are chemically coupled with an
underlying magma ocean.
However, a quantitative retrieval framework directly linking atmospheric spectra to
magma ocean properties has yet to be established. Here we introduce \textsc{MELTYQ}, a coupled magma-atmosphere retrieval framework that links transmission spectra to the oxidation state and volatile inventory of underlying magma oceans. \textsc{MELTYQ} combines a magma-atmosphere equilibrium model, which includes the solubility of H-/O-/C-/N-bearing species in the melt and redox reactions, with a Bayesian spectral retrieval scheme. Using simulated retrieval tests, we validate the approach and show that magma redox state and volatile content
can be constrained under favorable observational conditions. As a proof of concept, we apply \textsc{MELTYQ} to JWST transmission spectra of the benchmark sub-Neptunes K2-18\,b and TOI-270\,d.
We find that coupled magma-atmosphere retrievals are generally capable of reproducing the observed spectra of these planets. However, we identify several key limitations in the current framework. Specifically: more flexible free-retrieval approaches remain statistically preferred; the CO/CO$_2$ absorption feature near 4.5\,$\mu$m for TOI-270\,d is not fully captured; and a number of underlying model assumptions may not be strictly valid.
Nevertheless, embedding coupled magma-atmosphere models directly within Bayesian retrievals enables quantitative assessment of degeneracies and sensitivities, establishing a pathway for directly connecting atmospheric spectra to
magma composition in this underexplored exoplanet regime.
\end{abstract}

\section{Introduction} \label{sec:intro}
With the advent of high-precision atmospheric spectroscopy, the characterization of small exoplanets (i.e., sub-Neptunes and super-Earths) with radii barely larger than the Earth has recently made significant progress.
In particular, observations with the James Webb Space Telescope (JWST) have begun to provide unprecedented insights into the chemical diversity of sub-Neptunes, a category of small planets with lower densities than bare rocky planets. 
The transmission spectra of planets in this category, such as GJ~1214~b, K2-18~b, TOI-270~d, GJ 9827~d and GJ~3470~b, have revealed molecular signatures---including H$_2$O, CO$_2$, CH$_4$, SO$_2$, and other species \citep[e.g.,][]{Madhu_2023, Madhu_2025, Schlawin+24, Beatty+24, Holmberg_2024, Benneke+2024,Piaulet+24,Davenport+25,Hu_2025}---as well as featureless spectra, which suggests volatile enriched atmospheres or cloudy/hazy ones
\citep[e.g.,][]{Gao+23,Wallack+24,Gordon+2025}.
Future facilities such as Ariel and the ELTs will provide complementary capabilities to JWST for 100s of planets in this regime, further advancing our understanding of sub-Neptunes.\citep{Tinetti+2022,Dubey+2025}.

Sub-Neptune-sized planets are of particular interest for understanding planetary formation and evolution because they occupy a transitional regime between terrestrial planets and gas giants. However, their bulk compositions remain highly degenerate and cannot be uniquely constrained by mass--radius measurements alone \citep[e.g.,][]{Valencia+07, Adams+08}. These planets may consist of rocky interiors enveloped by thick hydrogen-dominated atmospheres, or alternatively may be rich in water and other volatiles. In the rocky-core scenario, theoretical models 
suggest that the high surface pressures ($\gtrsim 1\,\mathrm{Gpa}$) and temperatures ($\gtrsim 2000\,\mathrm{K}$) required to reproduce the observed mass and radius 
can lead to the formation of a long-lived or permanent magma ocean beneath the H$_2$-rich atmosphere \citep[e.g.,][]{Vazan+2018,Kite+2020}. Chemical interactions between such an atmosphere and an underlying magma ocean may significantly modify the atmospheric composition, thereby providing 
additional information
on the redox state and volatile inventory of the interior \citep[e.g.,][]{Kite+2020,Schlichting+2022,Charnoz+2023,Seo+2024,Tian+2024,Ito+2025,Werlen+2025,Bower+2025}. 

Within the rocky-core hypothesis (i.e., deep H$_2$ atmospheres), transmission spectroscopy only directly probes the upper atmospheric layers (i.e., $P \sim [1, 0.01]\,$bar). More specifically when the planetary mass is known, the absolute transit depth provides constraints on the total atmospheric mass, while the wavelength-dependent variations in the transit spectrum encode information about the chemical composition of this region. If the physical and thermochemical structure of the deeper atmospheric layers is sufficiently constrained, these observations can be used to infer the partial pressures of volatile species at the surface-atmosphere interface. Through dissolution equilibria and redox reactions, these partial pressures are directly linked to the abundances of volatile species dissolved in a putative magma ocean. As a result, atmospheric observations can in fact be leveraged to place constraints on both the volatile inventory and the redox state of the magma ocean. These parameters are key for constraining the bulk composition of the planetary interior and, by extension, the formation and evolutionary pathways of sub-Neptune planets. These pathways are shaped by the nature of the initial building blocks and by subsequent processes such as atmospheric escape and core differentiation \citep[e.g.,][]{Kite+2020,Bean+2021,Lichtenberg2021}.

Previous studies of sub-Neptune transit spectra typically employ two disconnected approaches: {\it i) atmospheric retrievals:} these are data-oriented approaches used to invert the observed spectrum and infer posterior distributions of the atmospheric composition. The model is statistically evaluated 100k--10M times; and {\it ii) forward melt-atmosphere coupling models:} these are magma+atmosphere forward models sparsely sampled (evaluated $\sim$ 100--1000 times) to reproduce the inferred composition from retrievals, or to match the observed spectra \citep{Shorttle+2024,Nixon+2025}. 
Although these two approaches are complementary and powerful for interpreting individual systems, they do not yet constitute a robust, integrated, and statistically rigorous framework. In particular, inferred interior properties are not consistently linked to observational uncertainties, and degeneracies within the parameter space cannot be fully quantified. As a consequence, a quantitative assessment of magma ocean properties and interior compositions of sub-Neptune planets based on transit spectroscopy has not yet been achieved.

In this paper, we propose that a direct specifically-designed retrieval framework can offer a promising path to addressing these limitations. As a proof of concept, we present the first coupled magma-atmosphere retrieval model explicitly designed to infer the volatile inventory and redox state of magma ocean in sub-Neptunes from transmission and emission spectra.
We focus on outlining the key assumptions of the model, demonstrating the scientific relevance of the approach, and assessing its computational feasibility using both simulated data and real observational cases.

The paper is organized as follows.
In Section~\ref{sec:method}, we describe the framework and its implementation. We validate the concept of our method using a simulated retrieval test in Section~\ref{sec:validation}. We then apply the retrieval model to the observed transmission spectra of the sub-Neptunes K2-18\,b and TOI-270\,d in Section~\ref{sec:application}. Finally, in Section~\ref{sec:discussion}, we discuss the limitations of our model and possible directions for improvement. This work is summarized in Section~\ref{sec:summary}.

\section{MELTYQ: A coupled retrieval framework for atmospheres and magma oceans}
\label{sec:method}
\begin{figure*}
\centering
    \includegraphics[width = 1.0\textwidth]{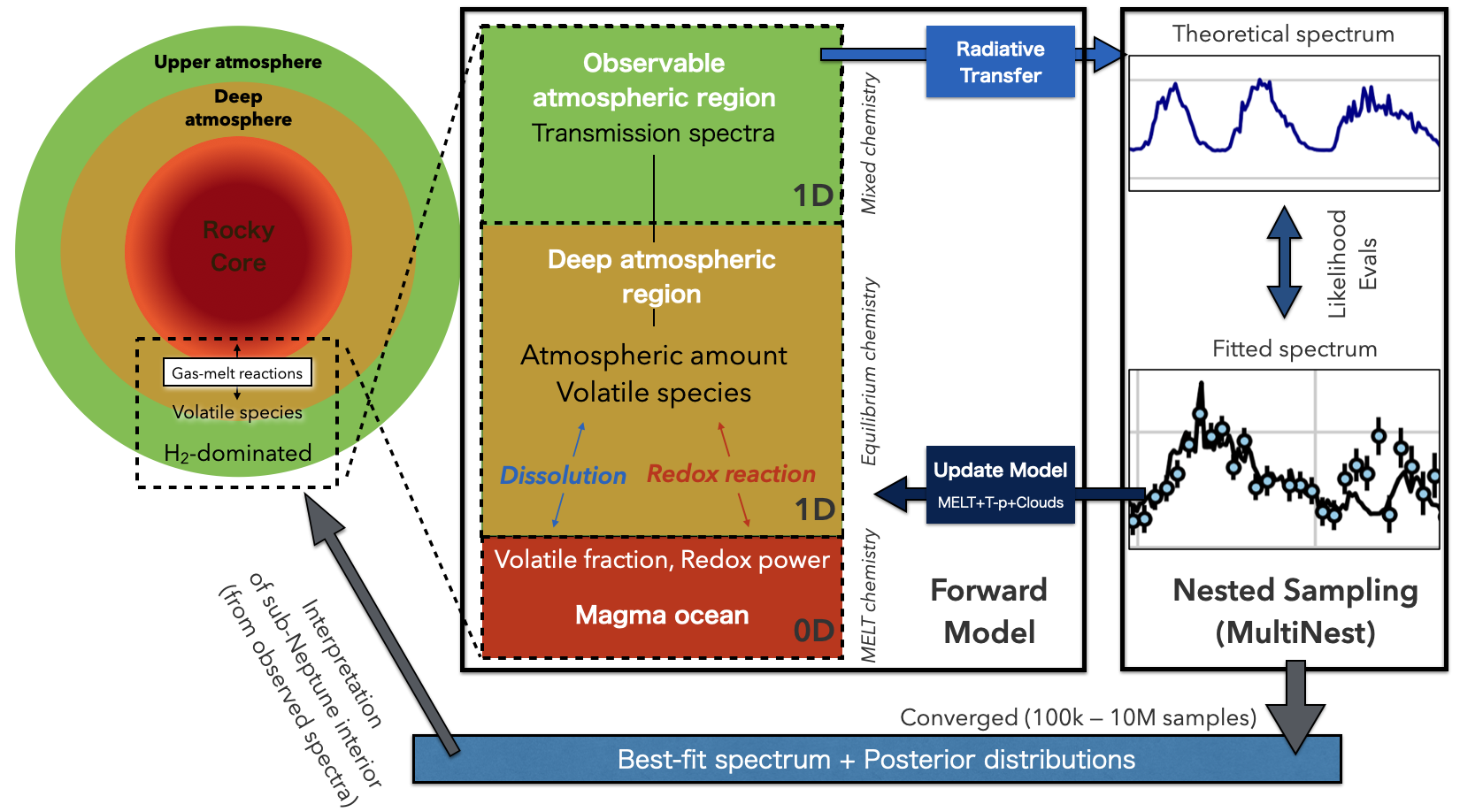}
    \caption{Schematics of the \textsc{MELTYQ} coupled magma-atmosphere retrieval. The model is valid for sub-Neptune hosting a thick H$_2$-rich atmosphere with enough surface temperature to melt rocks. The model splits the planet in three regions: a rocky core, a deep atmosphere, and an upper atmosphere. The chemistry is coupled allowing to retrieve the composition of the rocky magma via it's influence on the observable atmosphere.}\label{fig:concept}
\end{figure*}
For this demonstration, we develop a new forward model framework: \textsc{MELTYQ} (Magma-atmosphere Equilibrium and Layered ThermochemistrY retrieval-frameworQ). \textsc{MELTYQ} is integrated in the \textsc{TauREx} retrieval platform as a plugin \cite[version 3.1+, see:][]{Al-Refaie+2021, Al-Refaie+2022}. It uses the \textsc{TauREx} radiative transfer (RT) cores to compute transmission spectra from the low pressure atmospheric layers ($P \leq 10\,$bar set in this study), but uses a new description of sub-Neptune structure that couples the atmosphere with the underlying magma ocean (see conceptual diagram in Fig.~\ref{fig:concept}). Full Bayesian retrieval is automatically available, thanks to the \textsc{TauREx} integration. 

In \textsc{MELTYQ}, the planet is assumed to possess three regions: i) a rocky core with a magma ocean surface; ii) a deep atmosphere that is in direct contact with the magma surface; iii) an 
upper atmosphere probed by transmission spectroscopy.
Interfaces---from melt to deep atmosphere, and from deep to upper atmosphere---are defined by their pressure level: respectively $P_\mathrm{MELT}$ and $P_\mathrm{b}$. 
In the atmosphere, H$_2$, He, O$_2$, H$_2$O, CO, CO$_2$, CH$_4$, N$_2$ and NH$_3$ are considered. The partitioning of the H-/C-/O-/N-bearing species at the magma-atmosphere interface is governed by dissolution and redox reactions. The solubility laws are described in Sec~\ref{sec:magma} and summarized in Table~\ref{tab:soll}, while the reactions with given redox state of magma ocean are described in Sec~\ref{sec:deep} and summarized in R1--R4. 
For each region, we adopt simplified assumptions that are described below. The validity of these assumptions are also discussed in Sec~\ref{sec:discussion}, but we here focus on testing the viability of the concept and discussing potential challenges.

\subsection{Magma interior model}
\label{sec:magma}

For the rocky core, we treat the melt temperature ($T_\mathrm{MELT}$), pressure ($P_\mathrm{MELT}$), redox state, and volatile species fractions of the magma ocean surface. 
The redox state is expressed by the oxygen fugacity, $f_\mathrm{O_2}$, which quantifies how reduced or oxidized the melt is \citep[][and reference therein]{Frost+2008}. 
For the fractions of volatiles in melt, we consider H$_2$, H$_2$O, CO, CO$_2$, CH$_4$, and N$_2$, which are observed in experimental solubility measurements \citep[e.g.,][]{Holtz+2000,Hirschmann+2012,Yoshioka+2019,Dasgupta+2022}. We note that these parameters characterize the magma ocean surface in contact
with the atmosphere and we do not explicitly model the partitioning or reservoir sizes
of the entire magma ocean, mantle, or 
iron core, which is also discussed in Sec.~\ref{sec:interior}. In addition, we do not consider volatile partitioning into metal phases coexisting with silicate melt, which could be important if the differentiation of metallic iron and silicate is suppressed \citep[e.g.,][]{Lichtenberg2021,Young+2024}, as discussed in Sec.~\ref{sec:metal}.

We consider the solubility laws, which relate the abundance of each volatile species in the surface melt to its atmospheric species partial pressure (or fugacity), following previous magma--atmosphere coupling studies \citep[e.g.,][]{Kite+2020,Schlichting+2022,Tian+2024,Seo+2024}. For H$_2$O and CO$_2$, we adopt the empirical fits of \citet{lichtenberg+2021}, which were derived from laboratory solubility experiments \citep[see their Eq.~9 and Table~1]{lichtenberg+2021}. For H$_2$, CO, and CH$_4$, we use the formulations summarized in Eqs.~6, 15, and 17 of \citet{Seo+2024}, respectively, which are based on the experimental works of \citet{Hirschmann+2012}, \citet{Yoshioka+2019}, and \citet{Ardia+2013}. For N$_2$, we use the formulation given in Eq.~10 of \citet{Dasgupta+2022}.  Although the solubility law for N$_2$ depends on the logarithm of the oxygen fugacity relative to the iron-w\"{u}stite buffer ($\Delta \mathrm{IW}$) as well as the abundances of SiO$_2$, Al$_2$O$_3$, and TiO$_2$ in the melt, we focus solely on the effect of $\Delta \mathrm{IW}$ and assume basaltic melt compositions of SiO$_2$ (56\%), Al$_2$O$_3$ (11\%), and TiO$_2$ (1\%), following \citet{Dasgupta+2022}. While the rest of the melt composition is not explicitly assumed, it is expected to include the oxides of  Mg, Ca and Na, consistent with the basaltic compositions used in the calibration of the solubility law \citep[][]{Dasgupta+2022}. We compute $\Delta \mathrm{IW}$ from the assumed oxygen fugacity of the melt and the $(T,p)$-dependent function for the iron-w\"{u}stite buffer inferred from the experiments of \citet{Hirschmann2021}. 
The solubility laws adopted in our model, together with their experimental
calibration ranges in $T$ and $P$, are summarized in
Table~\ref{tab:soll}, and their functional forms are provided in
Appendix~\ref{app:solub}.

\begin{table}[b]
	\centering
	\caption{Solubility laws incorporated in MELTYQ}
 	\begin{tabular}{c c c} 
	Species & Experimental calibration ranges & Ref.  \\
    	 & in T [K] \& P [GPa] &   \\
    \hline
     H$_2$ & 1673--1773 \& 0.7--3 & A \\
     H$_2$O &  973--1723 \& $10^{-4}$--0.8  & B \\
     CO$_2$ & 1123--1923 \& $10^{-2}$--3 & B \\
     CO & 1523--1873 \& 0.2--3 & C \\
     CH$_4$ & 1673--1723 \& 0.7--3 & D \\
     N$_2$ & 1323--2600 \& $10^{-4}$--8.2 & E 
	\end{tabular}
\tablecomments{References are A, \citet{Hirschmann+2012}; B, \citet{lichtenberg+2021}; C, \citet{Yoshioka+2019}; D, \citet{Ardia+2013}; E, \citet{Dasgupta+2022}}
\label{tab:soll}
\end{table}

We also derive the radius of the rocky core, $R_\mathrm{core}$, from the planetary mass, $M_\mathrm{p}$, using the mass-radius relationship of \citet{Fortney+07}, assuming an innermost iron mass fraction equivalent to that of Earth’s core (33~\%) and neglecting the atmospheric mass contribution to the total planetary mass for simplicity. This assumption follows \citet{Rogers+2021}, which analyzed the properties and occurrence rates of planets with observed radii $R_\mathrm{p}\leq 4\,R_\oplus$ and orbital periods $P_\mathrm{orb}\leq 100$~days. In our retrievals, a planetary radius, $R_\mathrm{p}$ is not a free parameter, as opposed to standard atmospheric retrievals focusing only on the upper atmosphere. The radius solution, here defined as the radius at the bottom of the upper atmosphere region, is 
self-consistently calculated by integrating the hydrostatic equation (i.e., $dP/dr=-\rho GM_p/r^2$)
from $P_\mathrm{MELT}$ (i.e., from $R_\mathrm{core}$). During retrievals, the absolute {\it level} of the transmission spectrum strongly constrain the planetary radius, and hence $P_\mathrm{MELT}$. The details of the assumed atmospheric structure are described in Sec.~\ref{sec:deep} for $P \geq P_\mathrm{b}$ and in Sec.~\ref{sec:upper} for $P \leq P_\mathrm{b}$.
Although $R_\mathrm{core}$ can in principle be treated as a free parameter, we here chose to fix it, which provides more direct constraints to $P_\mathrm{MELT}$. In fact, $P_\mathrm{MELT}$ is also a proxy for the mass--radius solution, so it already provides important constraints on the radius solution, but note that this parameter also impacts the chemistry at the melt--atmosphere interface. 
Our assumptions and treatment of the rocky core radius in \textsc{MELTYQ} is further discussed in Sec.~\ref{sec:coreradi}.

\subsection{Deep atmosphere model}\label{sec:deep}

For the deep atmosphere, we set up a vertical one-dimensional atmospheric model in hydrostatic and chemical equilibrium, deriving the atmospheric properties between the upper boundary at $P_\mathrm{b}$ and the magma ocean surface at $P_\mathrm{MELT}$.
In this study, we set $P_\mathrm{b}=10$~bar, which is much deeper than the pressure level where the optical depth typically reaches unity in transmission or emission spectroscopy \citep[e.g.,][]{Heng+2017,Fortney+2019} and corresponds to where the chemical equilibrium assumption starts breaking due to vertical mixing \citep{Venot+2015, Al-Refaie+2024}. 
We treat the temperature at $P_\mathrm{b}$, labeled $T_b$, as a free parameter. 
For the $T-P$ profile in the deep atmosphere region, a simplified assumption is used. 
Because the region may not be governed solely by adiabatic processes due to Si-bearing species vaporized from magma ocean \citep{Misener+2022,Misener+2023}, and because observational constraints are limited, we approximate the temperature as varying linearly with $\log P$ between $T_\mathrm{MELT}$ at $P_\mathrm{MELT}$ and $T_\mathrm{b}$ at $P_\mathrm{b}$. Adopting this simplified profile is discussed as one of the caveats in \textsc{MELTYQ} in Sec.~\ref{sec:tp}.

We build the radial pressure--temperature--density profiles, and the chemical structure of the deep atmosphere accounting for the non-ideal behavior of the chemical species. 
We integrate the hydrostatic equation together with the simplified $T-P$ profile. 
We account for the altitude dependence of planetary gravity while neglecting the self-gravity of the atmosphere. 
The density, $\rho$, is computed using the additive-volume law, based on the local gas-species fractions and the corresponding equations of state (EOSs).
For H$_2$ and He, we use the EOS for an H--He mixture with a helium mass fraction of 0.275 from \citet{Chabrier+2021}. This is valid over a wide temperature range ($2 \le \log T\,\mathrm{[K]} \le 8$) and pressure range ($-5 \le \log P\,\mathrm{[bar]} \le 17$). 
As the resulting H$_2$ molar fraction of the H--He mixture is approximately 0.84 for this helium mass fraction, we adopt this fixed value and do not consider any depletion or enhancement of He, for consistency with the EOS, as also done in \citet{Ito+2025}.
For O$_2$, H$_2$O, CO, CO$_2$, and CH$_4$, we adopt the virial-type EOS of \citet{Zhang+2009}. \citet{Zhang+2009} derived this EOS using a virial-type expression based on a large number of  experimental data and molecular dynamics simulation data for temperatures of 673--2573~K and pressures of 0.001--10~GPa. 
We allow high-temperature extrapolation of the virial-type EOS but avoid extrapolation to higher pressures by imposing $P_\mathrm{MELT} \leq 10$~GPa, since virial expression could lose reliability at elevated densities (i.e., high pressures).
For N-bearing species, we neglect non-ideal effects and use the ideal EOS for simplicity, due to the lack of publicly available high-pressure EOS data up to 10~GPa.

For the chemistry, we also account for non-ideal effects of H-, C-, and O-bearing molecules by introducing the fugacity $f_s$ for each species $s$. 
The fugacity of species $s$ is defined as $f_s = \phi_s P_s$, where $\phi_s$ is the fugacity coefficient and $P_s$ is the partial pressure. 
For H$_2$, O$_2$, H$_2$O, CO, CO$_2$, and CH$_4$, we compute $\phi_s$ using the EOS of \citet{Zhang+2009}, whereas for N$_2$ and NH$_3$ we neglect non-ideal effects and set $\phi_s = 1$, for simplicity. We should note that neglecting non-ideal effects for nitrogen-bearing
species limits the quantitative accuracy of N$_2$ and NH$_3$ computed abundances, which is also discussed in Sec.~\ref{sec:eos}.
Fugacity coefficients are formally defined for pure components and not for mixtures. In this work, we approximate ideal mixtures using fugacity coefficients without accounting for mixture non-ideality \citep[e.g.,][]{Denbigh1981}, as also discussed in Sec.~\ref{sec:eos}.
Also, applying non-ideal corrections only to the atmospheric volatiles, while neglecting non-ideality in the silicate melt, may introduce systematic biases in the inferred atmospheric properties \citep[][]{Werlen+2026}. A fully self-consistent treatment including non-ideal effects in both atmospheric volatiles and melt phases is beyond the scope of this study and is left for future work.

For the gas chemistry in this region, chemical equilibrium is enforced assuming the following  reactions:
\begin{align}
\mathrm{O_2} + 2 \mathrm{H_2} \rightleftharpoons 2 \mathrm{H_2O}, \tag{R1}    
\label{eq:r1}
\\
2 \mathrm{CO} + \mathrm{O_2} \rightleftharpoons 2 \mathrm{CO_2}, \tag{R2}    
\label{eq:r2}  
\\
\mathrm{CO} + 3\mathrm{H_2} \rightleftharpoons  \mathrm{CH_4}+ \mathrm{H_2O}, \tag{R3}    
\label{eq:r3} 
\\
\mathrm{N_2} + 3\mathrm{H_2} \rightleftharpoons  2 \mathrm{NH_3}, \tag{R4} 
\label{eq:r4} 
\end{align}
These reactions describe the partitioning of H-, C-, O-, and N-bearing gas species, particularly  just above the magma ocean surface, where the partitioning is controlled by the redox state ($f_\mathrm{O_2}$) of the magma ocean and/or the total atmospheric hydrogen inventory.
These products and reactants in chemical equilibrium relate with each other, which is given by
\begin{eqnarray}
K=\prod (f_s/P_0)^a,
\label{eq:eq}
\end{eqnarray}
where $K$ is the chemical equilibrium constant, $a$ is the signed stoichiometric coefficient (positive for products, negative for reactants) and $P_0$ is the reference pressure of the
equilibrium constants, which is set at 1\,bar. The equilibrium constants are calculated using the differences in Gibbs free energy, taken from the JANAF database \citep{Chase1998}, between the reactants and the products. To obtain the chemical equilibrium composition between $P_\mathrm{MELT}$ and $P_\mathrm{b}$, we perform Gibbs free-energy minimization calculations along the temperature-pressure profile determined by $T_\mathrm{MELT}$-$P_\mathrm{MELT}$ and $T_\mathrm{b}$-$P_\mathrm{b}$ using a Newton-Raphson method and assuming vertically constant elemental abundances.

\subsection{Numerical procedure for magma-atmosphere coupling}\label{sec:numpro}
With the thermodynamic structure and chemical equilibrium relations of the
deep atmospheric region defined in Sec.~\ref{sec:deep}, we now summarize how the melt parameters determine the speciation and boundary conditions of the deep
atmosphere.
While we track H$_2$, H$_2$O, CO, CO$_2$, CH$_4$, and N$_2$ as volatile species in the silicate melt, as described in Sec.~\ref{sec:magma}, not all of these species are treated as free parameters. 
Their speciation is constrained by the solubility laws, the gas-phase equilibrium reactions, and the condition $P_\mathrm{MELT} = \sum_s P_{s,\mathrm{MELT}}$, where $P_{s,\mathrm{MELT}}$ is the partial pressure of species $s$ at the bottom of the deep atmosphere. 

In \textsc{MELTYQ}, we therefore treat only the nitrogen abundance in the melt ($N_\mathrm{MELT}$), the carbon monoxide abundance in the melt ($\mathrm{CO}_\mathrm{MELT}$), the oxygen fugacity ($f_{\mathrm{O}_2}$), the melt-surface pressure ($P_\mathrm{MELT}$), and temperature ($T_\mathrm{MELT}$) as free parameters; the remaining volatile fractions in the melt are derived quantities.
At $P_\mathrm{MELT}$, this system contains 12 unknowns: eight partial pressures (or fugacities) of H$_2$, He, H$_2$O, CO, CO$_2$, CH$_4$, N$_2$, and NH$_3$, and four fractions of H$_2$, H$_2$O, CO$_2$, and CH$_4$ in melt. 
These are solved using 12 equations: the six solubility relations summarized in Table~\ref{tab:soll}, the four gas-gas equilibrium relations R1-R4, the fixed H$_2$ molar fraction of the H--He mixture, and the closure condition $P_\mathrm{MELT} = \sum_s P_{s,\mathrm{MELT}}$, for a given set of free parameters ($N_\mathrm{MELT}$, $\mathrm{CO}_\mathrm{MELT}$, $f_{\mathrm{O}_2}$, $P_\mathrm{MELT}$, and $T_\mathrm{MELT}$). 

This choice of free parameters is motivated by the goal of describing the magma-atmosphere interface using a minimal and physically interpretable set of variables. The parameters $f_{\mathrm{O}2}$ and $T_\mathrm{MELT}$ characterize the redox and thermal state of the surface magma, respectively, while $P_\mathrm{MELT}$ sets the pressure at the interface and controls the total atmospheric mass.
For volatile species, $N_\mathrm{MELT}$ directly represents the nitrogen abundance in the melt. For carbon, although the total carbon abundance is the physically relevant quantity, it is determined implicitly in our model. We therefore adopt $\mathrm{CO}_\mathrm{MELT}$ as a representative parameter, noting that alternative choices such as $\mathrm{CO}_{2,\mathrm{MELT}}$ or $\mathrm{CH}_{4,\mathrm{MELT}}$ would provide equivalent information due to their interconversion through redox reactions.

Using the resulting gas species fraction at $P_\mathrm{MELT}$ as the lower boundary condition, the thermodynamic structure and chemical equilibrium relations of the deep atmosphere are integrated upward from $P_\mathrm{MELT}$ to
$P_\mathrm{b}$.
Therefore, for a given set of melt free parameters ($N_\mathrm{MELT}$, $\mathrm{CO}_\mathrm{MELT}$, $f_{\mathrm{O}_2}$, $P_\mathrm{MELT}$, and $T_\mathrm{MELT}$) and the temperature at the upper boundary interface ($T_b$), the model of the deep atmospheric region derives the vertical distribution of gas species in hydrostatic and
chemical equilibrium with the underlying magma ocean surface. 
A numerical test of our magma--atmosphere coupling simulation code against \citet{Seo+2024} is presented
in Appendix~\ref{app:bench}.
The parameter dependence of all derived quantities at $P_\mathrm{MELT}$ and
$P_\mathrm{b}$ is shown in Appendix~\ref{app:para}.

\subsection{Upper atmosphere model}
\label{sec:upper}
For the upper atmospheric region with $P \le P_\mathrm{b}$ and where the radiative transfer will be performed, we assume hydrostatic equilibrium but do not impose chemical equilibrium. We model a plane-parallel atmosphere with 100 log-spaced pressure layers from $P \in [10^6, 10^{-5}]\,$Pa.
At these pressures, non-equilibrium processes such as vertical mixing and photochemistry can dominate, making chemical equilibrium inappropriate \citep{Venot+2015, Al-Refaie+2024}. 
Although incorporating a full non-equilibrium chemistry model such as \textsc{FRECKLL} \citep{Al-Refaie+2024} would be possible in the future, as also discussed in Sec.~\ref{sec:eq}, we here adopt a vertically uniform atmospheric composition for simplicity. This assumption remains standard in the field and is sufficient for the purpose of this demonstration. 
Non-ideal effects are also expected to be negligible in this low-pressure region \citep[e.g.,][]{Tian+2024}, and we therefore assume an ideal gas law.
Here, the observed spectrum probe variations in the temperature profile of \cite{Schleich_2024}, so we need to consider non-isothermal behavior. This is done by retrieving three additional $T-P$ points at fixed pressures, $P \in \{10^4, 100, 10^{-2}\}$\,Pa, where the temperature is linearly interpolated. Above $P =10^{-2}\,$Pa, the temperature profile is considered isothermal. 

Clouds and hazes can also be present in sub-Neptunes atmospheres, and will impact the observed spectrum. To model these, we choose a free approach. This is justified by our lack of prior knowledge on cloud formation and properties in this class of planets. We include two commonly used components: i) a cloud deck characterized by the top pressure ($P_\mathrm{CLOUDS}$) bellow which the atmosphere is fully opaque; and ii) a wavelength dependent haze layer following the formalism from \cite{Lee_2013}. This latter phenomenological model is parameterized by the aerosol particle radius ($R_\mathrm{LEE}$) in $\mu$m, the Efficiency coefficient ($Q_\mathrm{LEE}$), the particle number density ($X_\mathrm{LEE}$) in particles/m$^3$, the pressure level of the cloud layer ($P_{LEE}$), and the extent of the cloud layer \cite[in this study, fixed to 2 in pressure log-space since it's degenerate with the other parameters][]{Changeat+2025}.
The radiative transfer is performed in this region only using the base transmission RT core from the \textsc{TauREx} plateform. The RT calculations include clouds and hazes (see above), molecular absorption from H$_2$O \citep{polyansky_h2o}, CO \citep{li_co_2015}, CO$_2$ \citep{Yurchenko_2020}, CH$_4$ \citep{ExoMol_CH4_new}, NH$_3$ \citep{Yurchenko_2011_NH3}, using the \textsc{ExoMol} line-lists at resolution $\mathcal{R} = 50,000$ \citep{Tennyson_exomol, Chubb_2021_exomol}, collision induced absorption CIA \citep{abel_h2-h2, abel_h2-he} from HITRAN, and Rayleigh Scattering \citep{cox_allen_rayleigh}. In one scenario for TOI-270\,d, we also retrieve CS$_2$ \citep{Sharpe_2004}, which is extracted from the HITRAN database \citep{GORDON_2026}.

\subsection{Bayesian optimization}

The full forward model includes free parameters (to be retrieved) from each region: 
\begin{itemize}
    \item Magma interior: $f_{O_2}$, CO$_\mathrm{MELT}$, N$_\mathrm{MELT}$;
    \item Deep atmosphere: $T_\mathrm{MELT}$, $P_\mathrm{MELT}$, $T_\mathrm{b}$;
    \item Upper (observable) atmosphere: $T_1$, $T_2$, $T_3$, $P_\mathrm{CLOUDS}$, $R_\mathrm{LEE}$, $Q_\mathrm{LEE}$, $X_\mathrm{LEE}$, and $P_{LEE}$.
\end{itemize}

Note that all the free chemical parameters in this retrieval model refer to the magma composition. This means that retrieving observed spectra with this coupled model directly infer the interior composition of the planet. We explore the free parameter space using uninformative priors (i.e., large uniform priors) via the Nested Sampling Bayesian optimizer \textsc{MultiNest} \citep{Feroz+2009, Buchner+2016}, already made available in the \textsc{TauREx} framework. We use 1000 live points and an evidence tolerance of 0.5.
When combining dataset from different instruments (e.g., NIRISS, NIRSpec, NIRCam, and MIRI), we also retrieve a wavelength-independent offset for each dataset \cite[see discussion and justification in][]{Edwards+2024}.

\begin{figure}
\centering
    \includegraphics[width = 0.5\textwidth]{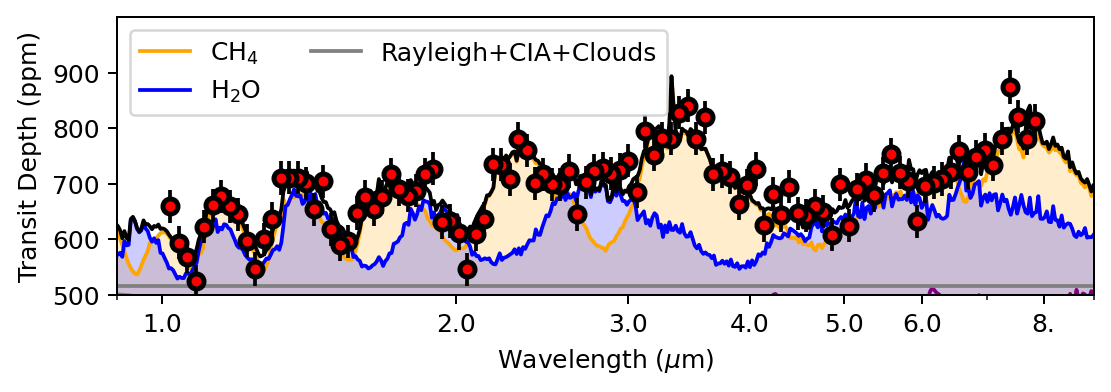}
    \includegraphics[width = 0.45\textwidth]{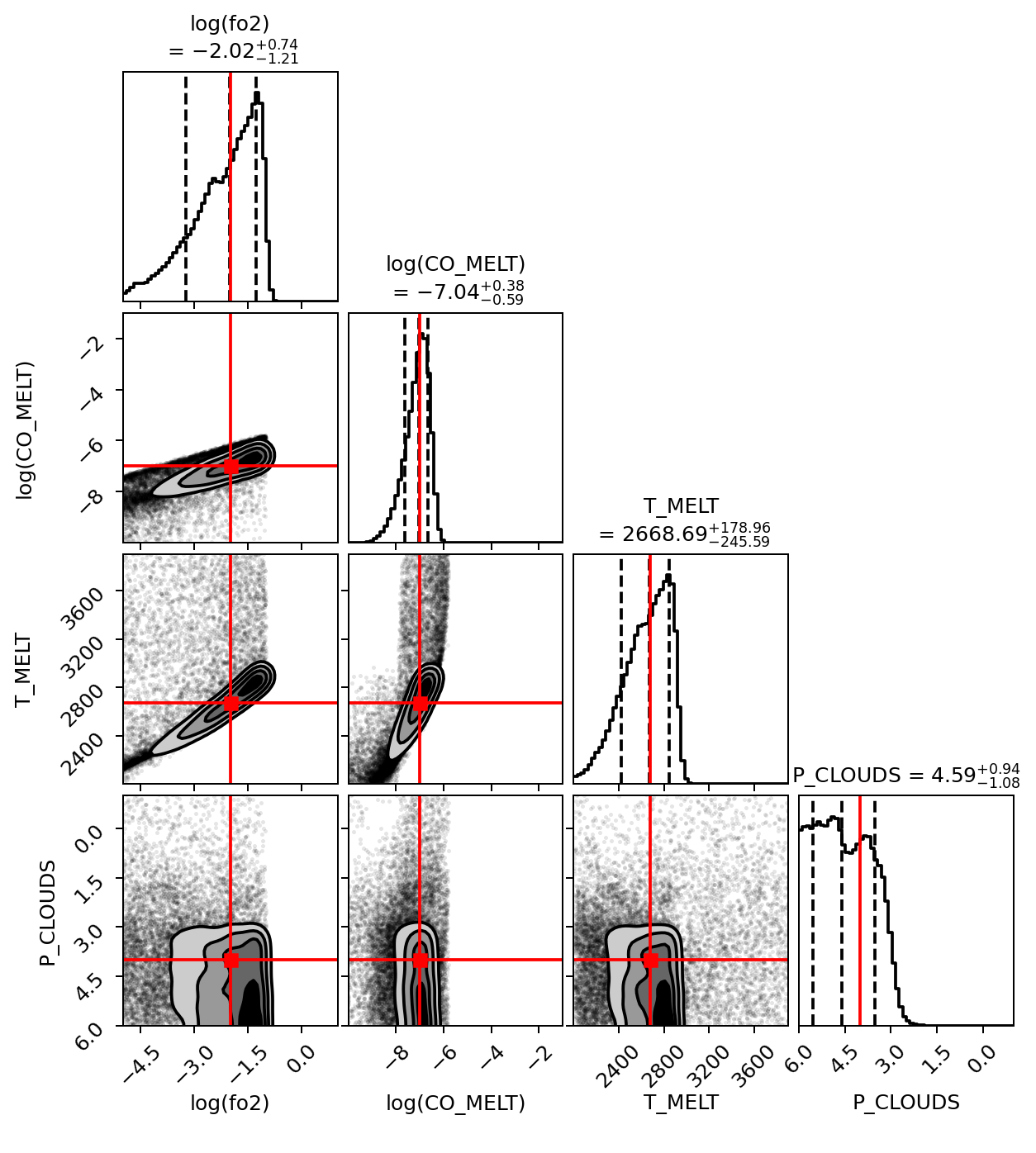}
    \caption{Synthetic retrieval using the MELTYQ lava-ocean--atmosphere framework. Top: simulated spectrum with the retrieval fit; Bottom: posterior distribution with ground-truth (red crosses). This test shows the feasibility of retrieving the magma composition of sub-Neptunes from atmospheric spectra.}\label{fig:simu_retrieval}
\end{figure}

\begin{figure*}
\centering
    \includegraphics[width = 0.99\textwidth]{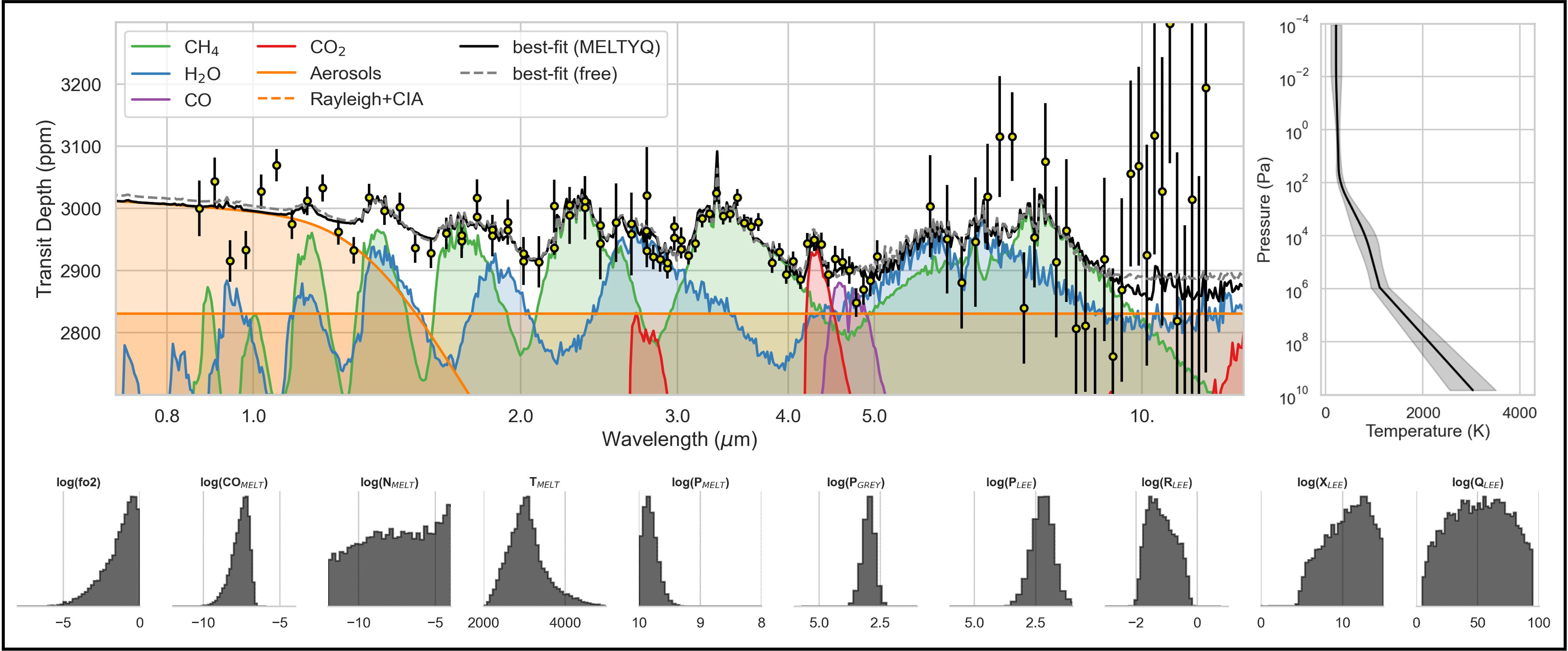}
    \caption{Retrieval results for K2-18\,b. \textbf{Top left}: observed spectra with best-fit MELTYQ retrieval (solid black) and free retrieval (dashed grey). Contribution functions for CH$_4$, H$_2$O and continuum opacities (Rayleigh scattering, CIA, and aerosols) are also shown. \textbf{Top right}: recovered T--p profile for the MELTYQ run. \textbf{Bottom}: corresponding posterior distributions. Note that these retrievals do not necessarily constitute the best achievable fits of the data but serve to illustrate the relevance of our coupled magma-atmosphere approach. }\label{fig:spectra}
\end{figure*}

\section{Validation of the magma retrieval concept}
\label{sec:validation}
To validate our \textsc{MELTYQ} coupled-magma implementation, we first perform a self-retrieval test \cite[i.e., the retrieval model has the same assumptions as the forward model simulation, see similar procedures in e.g.,][]{Schleich_2024, Changeat_2024_explor}. We produce a synthetic spectrum simulation and retrieve the free chemical parameters using the same forward model. We briefly summarize the main features of the simulation here: a synthetic K2-18\,b-like exoplanet (i.e., same bulk parameters), assuming a deep magma ocean at $P_\mathrm{MELT} = 10$\,GPa and temperature $T_\mathrm{MELT} = 2675$\,K. The magma composition is set at $f_{\mathrm{O_2}} = 0.01$\,Pa, $\mathrm{CO_{MELT}} = 10^{-7}$, and $\mathrm{N_{MELT}} = 10^{-4}$. These values are chosen to produce a mixed H$_2$/CH$_4$/H$_2$O atmosphere. We assume an isothermal $T-P$ profile at $T=675\,$K for $P\geq10$\,bar. High pressure Grey clouds are also added at $P = 10^4$\,Pa, but we do not consider wavelength-dependent cloud opacity for this simplified example. The high-resolution spectrum is then binned down to resolution $\mathcal{R} = 50$ between $\lambda \in [1, 7.8]\,\mu$m\footnote{for this demonstration, this is inspired from the ESA-Ariel mission wavelength regime, but does not need to be.}. White gaussian noise at 30\,ppm level is finally added to produce our simulated observation.

The simulated observations and the self-retrieval results are shown in Figure \ref{fig:simu_retrieval}. As can be seen in the posterior distributions, the differences in the magma composition are traceable to the atmosphere, allowing for a retrieval to theoretically constrain magma composition from spectra. We have also tested the same self-retrieval but with unscattered simulated observations. As expected, we find that the retrieved parameters match the inputs parameters even more closely. These tests validate our code integration and the relevance of our approach. From a computational point-of-view, \textsc{MELTYQ} is highly optimized with constant convergence times of $\lesssim 5\,$s across the valid parameter space. This implied that retrievals can be done in $\sim 1$\,day on modern HPC. For reference, the average computational requirement for a typical MELTYQ forward model---including radiative transfer calculations---range from $\sim 1.1\,$s in the TOI-270\,d case, to $\sim 3.2\,$s for the K2-18\,b case.

\section{Application to a real JWST observation}
\label{sec:application}

\subsection{K2-18\,b}

We applied our MELTYQ retrieval approach to two available JWST spectra of sub-Neptunes falling in the valid physical regimes of our model: K2-18\,b, and TOI-270\,d. The reduced spectra were obtained from \cite{Madhu_2023, Hu_2025, Madhu_2025} for K2-18\,b (respectively, NIRISS, NS-G235H + NS-G395H, and MIRI-LRS), and from \cite{Holmberg_2024} for TOI-270\,d (NS-G395H), via their associated OSF repositories (see Data Availability Statement). K2-18\,b and TOI-270\,b have been described as potential Hycean planets. Their spectra are consistent with mainly spectral modulation from CH$_4$ and presence of some clouds/hazes \citep{Liu_2025}. There are evidence for additional species, e.g., CO$_2$ and H$_2$O, but abundances and significance of the detections vary depending on dataset and studies \citep{Madhu_2025, Hu_2025}. We first concentrate on the K2-18\,b case, which has the most extensive dataset of these planets. For K2-18\,b, we conduct two simple atmospheric retrieval scenarios: 
\begin{enumerate}
    \item {\it MELTYQ retrieval} (S1): this uses the newly coupled magma--atmosphere model. Chemistry related free parameters relate to the composition of the magma only. The planetary radius is not retrieved but calculated self-consistently.
    \item {\it Free retrieval} (S2): a standard free chemistry retrieval with non-informative priors as baseline. The planetary radius is retrieved.
\end{enumerate}

The results for K2-18\,b are summarized in Figure \ref{fig:spectra}, while the full posterior distributions can be found in Figure \ref{fig:corner_k218}. As can be seen, the spectrum of K2-18\,b---as observed from JWST---leads to constraints on their interiors when using a \textsc{MELTYQ} retrieval. 

\begin{figure}
\centering
    \includegraphics[width = 0.49\textwidth]{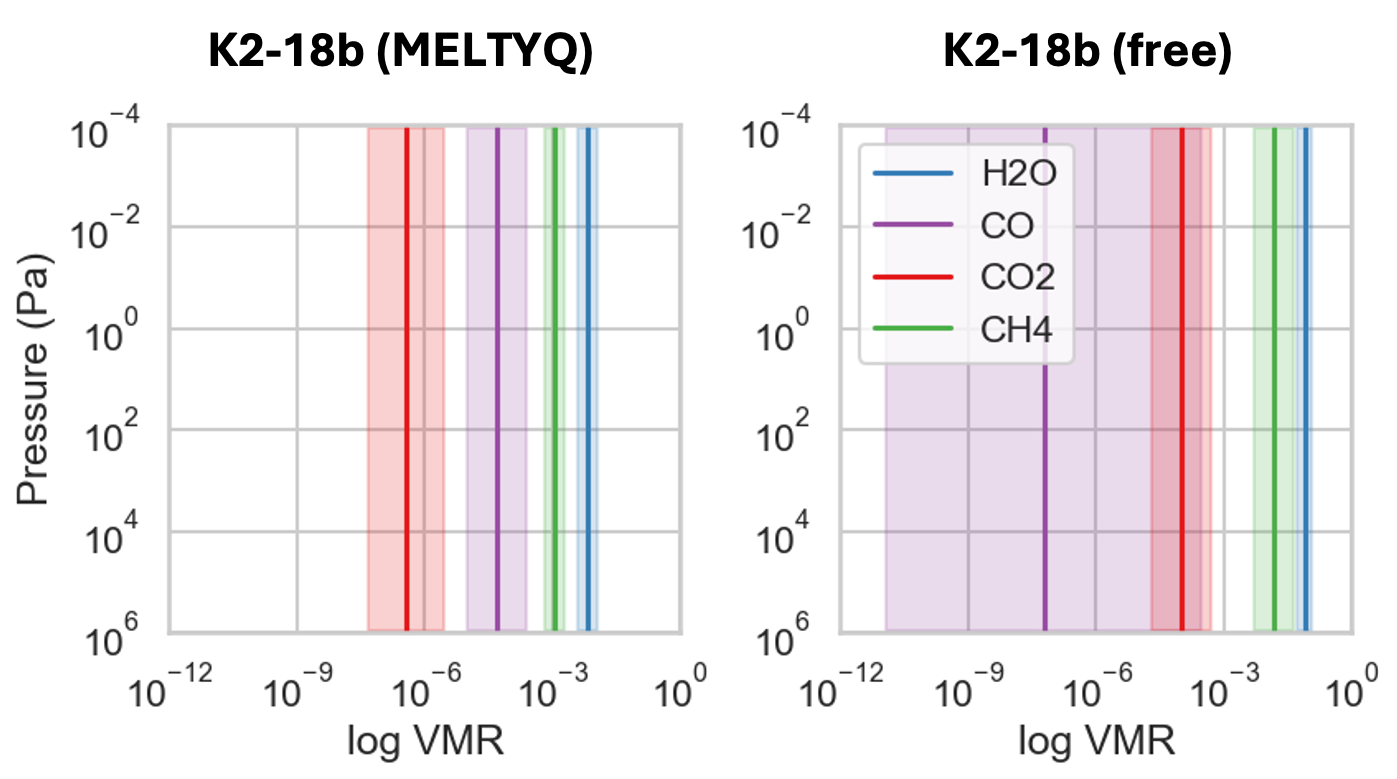}
    \caption{Retrieved chemistry for K2-18\,b. While qualitatively similar, the abundances retrieved by the MELTYQ and free retrievals showcasing the model dependence of retrievals for small exoplanets.}\label{fig:chemistry}
\end{figure}

The data suggest that a deep $\sim$7\,GPa atmosphere with high magma temperatures ($\sim$3000\,K) could explain the K2-18\,b spectrum. The magma--atmosphere interface is constrained because we utilize self-consistent assumptions for the planet's structure---i.e., the planetary radius is not directly retrieved---and we use a fixed planetary mass from radial velocity estimates. It is also found that a high value of $f_{\mathrm{O_2}}$ is preferred, with values ranging from log($f_{\mathrm{O_2}}) \in [-4; 1]$\,Pa. A {\it linear} correlation exists between $f_{\mathrm{O_2}}$ and $\mathrm{CO_{MELT}}$, as explicited in Figure \ref{fig:corner_k218}. This is because CH$_4$ is the most strongly visible absorber in K2-18\,b, which depends on both $f_{\mathrm{O_2}}$ and $\mathrm{CO_{MELT}}$. 

With both of our approaches, we find a qualitatively similar interpretation to \cite{Liu_2025}, which used free chemistry. The solution suggests that clouds or hazes could strongly affect the data continuum---especially at short wavelengths---and lead to lower atmospheric fractions of H$_2$O and CH$_4$. In both runs, H$_2$O and CH$_4$ are the dominant trace species, with volume mixing ratios ranging from 0.1--1\% (see Figure \ref{fig:chemistry}).
Note that, in addition to the effects of clouds and hazes on spectra, the condensation of H$_2$O may further reduce its gas-phase abundance, particularly in cooler atmospheric regions. Since condensation processes are not included in the present model, the inferred H$_2$O abundances may be overestimated where condensation is efficient.
At the same time, if clouds are composed of H$_2$O, they may obscure deeper atmospheric layers and lead to an underestimation of H$_2$O abundances below the cloud deck. A self-consistent retrieval including condensation and cloud formation \citep[e.g.,][]{Ma+2023} would be required to reduce such bias by properly accounting for these effects, but this is beyond the scope of this study.

Comparing the log evidence ($ln\,\mathcal{E}$), we note an overall preference for S2 over S1: $ln\,\mathcal{E}_{\mathrm{free}} =907.9$ vs $ln\,\mathcal{E}_{\mathrm{MELTYQ}} = 901.2$. From a pure Bayesian model comparison standpoint, this makes sense given the additional degrees of freedom in the free retrieval. Therefore, self-consistent MELTYQ models are not favored compared to the free retrievals. The $ln\,\mathcal{E}$ penalty, however, is comparable in magnitude to what is typically obtained by chemical equilibrium schemes versus free retrievals \citep{Changeat_2022_fiveKey}.

\subsection{TOI-270\,d}

\begin{figure}
\centering
    \includegraphics[width = 0.49\textwidth]{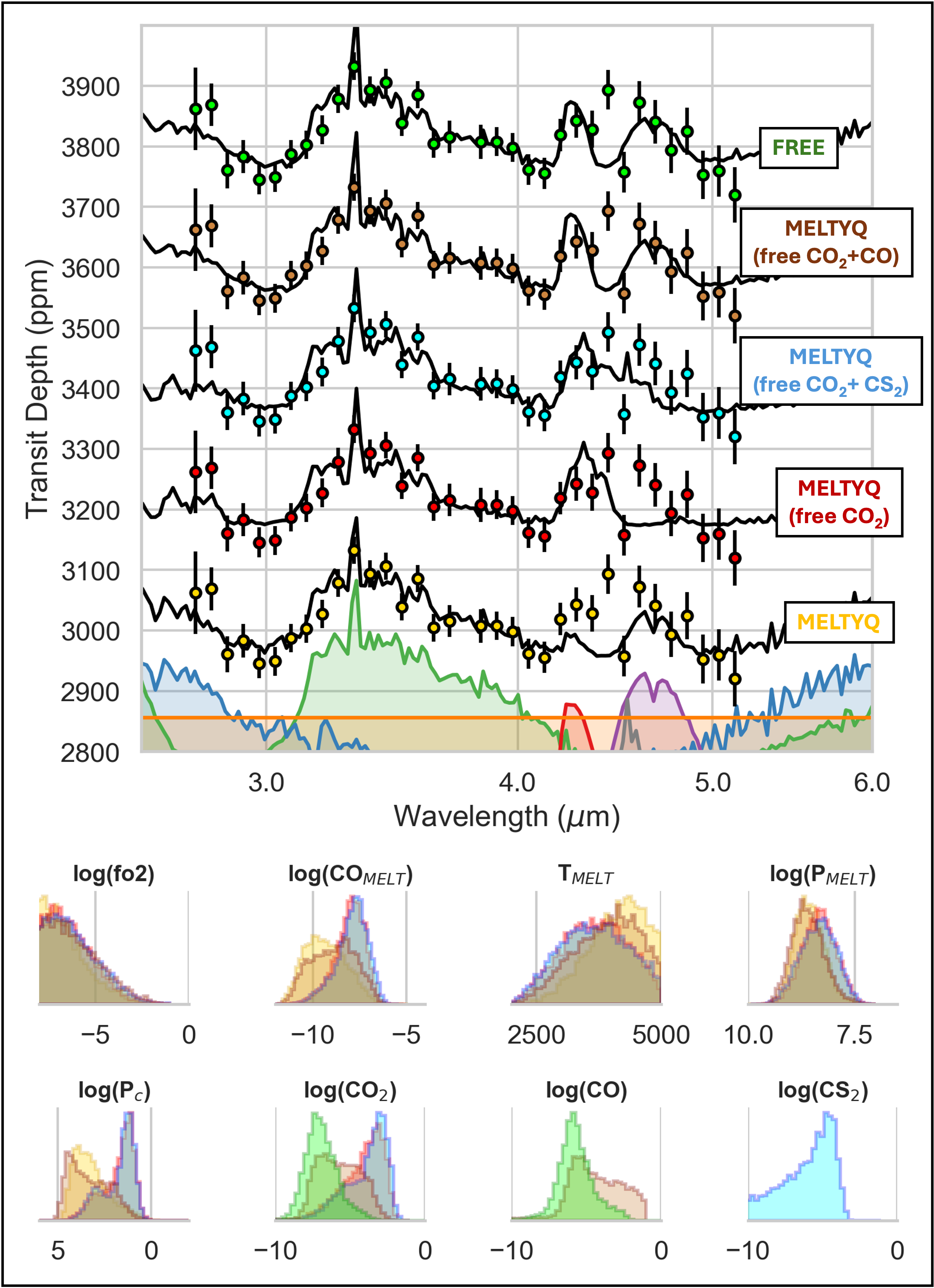}
    \caption{Retrieved best-fit models for TOI-270\,d (top) and posterior distributions (bottom). The 4.5$\mu$m spectral feature is not easily explained by the MELTYQ retrievals, and more flexible hybrid or free approaches are needed. When leaving CO, CS$_2$, and CO$_2$ as free parameters, the spectrum can be well explained, suggesting that TOI-270\,d breaks the assumptions of MELTYQ (i.e., no lava ocean, important dis-equilibrium processes, etc).}\label{fig:combined_toi}
\end{figure}

For TOI-270\,d, the available data does not span a large wavelength range. The available reduced data for TOI-270\,d covers $\lambda \in [2.6; 5]\,\mu$m, clearly probing H$_2$O, CH$_4$, CO, and CO$_2$ as seen in the high-quality spectrum, but the lack of baseline does not allow to anchor the continuum accurately. Additional data from NIRISS exist for TOI-270\,d, but the final data products are not available at present. The NIRSpec data is consistent with a strong CH$_4$ feature (see Figure \ref{fig:combined_toi}). Given the magnitude of this feature, a lighter atmosphere than K2-18\,b is needed and the MELTYQ retrieval favors a lower $f_{\mathrm{O_2}}$. The pressure interface is also found to be less deep, at $P_\mathrm{MELT} \sim  0.3$\,GPa, and consistent across all the models. These differences lead to a CH$_4$ VMR of about 10$^{-4}$ for TOI-270\,d. The spectral signal around $\lambda = 4.5\,\mu$m is not well explained by the pure MELTYQ model (see Figure \ref{fig:combined_toi}). Most likely, this is because the CH$_4$, CO$_2$ and CO are not in equilibrium balance at 10 bar, which is set as the boundary between the chemically equilibrated deep atmospheric region and the uniform upper atmospheric region, or because other species (such as CS$_2$), which are not included in MELTYQ but may play a significant role in this atmosphere, contribute to the observed discrepancy. With MELTYQ, an hybrid approach is possible: individual molecules can be decoupled from the self-consistently computed chemistry. This is relevant here because MELTYQ---and any other self-consistent chemical model---could miss important {\it dis-equilibrium} processes. To test this, three additional scenarios are performed: S3 where CO$_2$ is decoupled; S4 where both CO$_2$+CS$_2$ are decoupled; and S5 where both CO+CO$_2$ are decoupled. We obtain much better fits with these scenarios. For instance, we find $ln\,\mathcal{E}_{\mathrm{free}} = 288.8$ to be compared with $ln\,\mathcal{E}_{\mathrm{S4}} =285.8$ and $ln\,\mathcal{E}_{\mathrm{S5}} =287.0$. S5 actually nearly reproduces the free retrieval. Our results show that interpretation of the CO/CO$_2$ band is highly sensitive to model choice and is largely driven by two widely separated datapoints near the center of the absorption feature, which none of the models fully reproduce. The inferred behavior in this region depends strongly on the balance between model flexibility and individual datapoints, making it prone to over-interpretation---particularly in the presence of instrumental systematics or statistical noise. Notably, removing a single datapoint would yield an excellent fit. More generally, our exploration demonstrates that a hybrid approach enables the treatment of chemically dis-equilibrated compositions in retrievals of transitional atmospheres. 
Looking at the distributions in Figure \ref{fig:combined_toi}, the interpretation of TOI-270\,d's magma composition seems independent from the hybrid treatment and we always recover a low $f_{\mathrm{O_2}}$ and $\mathrm{CO_{MELT}}$ (log($f_{\mathrm{O_2}}) < -4.0$ and log($\mathrm{CO_{MELT}}$)\,$\in [-10.5, -7.0 ]$).

\section{Discussion: physically valid regimes and caveats}\label{sec:discussion}

Our framework is constructed under the conceptual assumption that the atmospheric
composition is strongly coupled to the underlying magma ocean through the 
dissolution and redox reactions of volatile species (Fig.~\ref{fig:concept}). 
Because of this, the framework is not appropriate for sub-Neptunes 
with icy or miscible interiors. The latter conditions may arise even in the rocky-core scenario. 
Furthermore, we assume that atmospheric mixing is sufficient to maintain vertically constant elemental abundances,  
that may not hold in all cases.
Here, we discuss the regimes that are physically inaccessible to our framework in its current state.

\subsection{High pressure fully miscible interiors}
\label{sec:misc}
At sufficiently high pressures and temperatures, hydrogen and silicate melts
can become fully miscible, eliminating a sharp distinction between the rocky
interior and the overlying H$_2$ atmosphere in sub-Neptunes
\citep[e.g.,][]{Markham+2022,Young+2025,Rogers+2025}.
In such a fully miscible regime, a distinct silicate dominated magma-atmosphere
interface does not exist. Solubility
laws and redox equilibria at a
silicate dominated melt surface are therefore no longer meaningful.

Recent first-principles calculations predict that hydrogen--silicate miscibility
is reached at temperatures of order
$T \sim 3000$--$4000\,\mathrm{K}$
and pressures of order
$P \sim 1$--$10\,\mathrm{GPa}$ or higher, depending on composition
\citep{Young+2024,Young+2025,Rogers+2025,Gilmore+2026}.
Under these conditions, the interior and the envelope may become miscible at high pressures and temperatures, while remaining phase-separated at lower pressures and temperatures corresponding to the surface. The transition between these regimes is defined by the binodal (or solvus), which is the thermodynamic stability boundary in pressure-temperature space separating single-phase and two-phase regions for silicate-hydrogen mixtures \citep{Young+2024}.
Because our framework relies on the existence of a chemically coupled but
physically distinct silicate dominated magma ocean surface in contact with an atmosphere (i.e., a system operating below the binodal), it
cannot be applied to planets residing in this fully miscible regime. In our retrieval framework, these regimes can be explicitly rejected using the \textsc{TauREx} exception system to assign infinitesimally low likelihoods. In this paper, we apply this rejection mechanism for $P_\mathrm{MELT} > 10\,$GPa. For K2-18\,b, we advise caution when interpreting our results: the retrieved solution is near the potential miscible regimes, so the assumptions underlying the model may no longer be satisfied. Based on the critical T-P conditions reported by \citet{Gilmore+2026}, using a linear approximation\footnote{We approximate the critical temperature as a linear function of pressure, using the critical T-P conditions (3,500 K at 2 GPa and 2,600 K at 10 GPa) shown in \citet{Gilmore+2026}.}, we estimate that $\sim$40\% of the posterior samples fall within the regime where miscibility may occur.

\subsection{Stratified atmospheres: poor vertical mixing}
\label{sec:mix}
Our framework implicitly assumes that the elemental composition inferred
from the observable atmosphere reflects the composition of the deeper
atmospheric layers that are in contact with the magma ocean.
However, condensation of refractory species near the magma ocean surface
can generate mean-molecular-weight gradients in the deeper atmosphere
\citep{Misener+2022,Misener+2023}.
Once established, such gradients in hydrogen-dominated atmospheres can
strongly suppress large-scale convection and vertical mixing, as
demonstrated in the context of water condensation
\citep[e.g.,][]{Leconte+2017,Leconte+2024,Habib+2025}.
If the resulting stratified layer extends over a thickness of
$\sim0.1\,R_\oplus$ \citep{Misener+2023}, and the reduced eddy diffusivity
within this layer falls in the range $10^{-4}$--$10^{1}\,\mathrm{m^2\,s^{-1}}$,
as inferred from atmospheric dynamics simulations of water condensation-induced
gradient \citep{Leconte+2024,Habib+2025}, the corresponding vertical
mixing timescale spans $\sim10^{-3}$--$10^{2}$~Myr.
Thus, if vertical mixing is sufficiently slow, the deep atmospheric part may become chemically isolated from the observable region, preventing interior signatures from being transmitted to the upper atmosphere.
Under these conditions, our framework assumptions are not valid and cannot be used directly.

\subsection{Interior beneath the surface magma ocean}\label{sec:interior}
The constraints provided by \textsc{MELTYQ}  primarily apply to the
magma ocean surface in contact with the atmosphere.
While the retrieved redox state and volatile abundances provide important
information about the near-surface melt, they do not directly probe the
composition of the deeper mantle or the metallic iron core.
Any influence of the deep interior on the observable atmosphere is therefore
necessarily indirect, operating through processes such as redox buffering
and the exchange of volatiles between the surface melt and the atmosphere.
As a result, the constraints obtained with \textsc{MELTYQ} should in fact be
interpreted as applying to the magma ocean surface rather than to the bulk
mantle or core composition. Extending the framework to more explicitly connect this surface layer to the
deeper mantle and core, for example through models of melt
convection, differentiation, and core-mantle equilibration, represents a natural direction for future work.

\subsection{Undifferentiated metallic iron in magma}\label{sec:metal}
Whether sub-Neptunes with rocky interiors possess differentiated iron-dominated cores remains an open question \citep{Lichtenberg2021,Young+2024,Young+2025}.
A recent study based on phase equilibria in the MgSiO$_3$-Fe-H$_2$ system suggests that silicate and metal phases may become largely miscible under core conditions in sub-Neptunes, while they can remain separated in shallower regions \citep{Young+2025}.
In addition, even if metal and silicate are chemically separated, a highly turbulent magma ocean and the density deficit of metal containing volatiles may suppress efficient metal-silicate differentiation in sub-Neptunes \citep{Lichtenberg2021,Young+2024}.
If the interior is undifferentiated, as suggested by these studies, metal coexisting with silicate melt may act as an important reservoir for volatile species, potentially affecting the atmospheric composition \citep[e.g.,][]{Schlichting+2022,Werlen+2025}.
In this work, we do not consider volatile partitioning into metal phases, which could be important in such undifferentiated cases. Instead, MELTYQ constrains the surface silicate melt and does not explicitly account for metal phases contributions, as also discussed in Sec.~\ref{sec:interior}. Incorporating the effects is beyond the scope of this study, but could be explored as a possible direction for future improvements of MELTYQ.

\subsection{Uncertainty in the core radius}
\label{sec:coreradi}
In this study, we do not treat the rocky core radius as a free parameter.
Instead, for a given planetary mass, we assume a representative core composition
corresponding to an Earth-like iron mass fraction, and retrieve the surface pressure at the atmosphere--interior interface. This retrieved parameter acts in a similar way to the observed radius \citep[typically at $P = 10$\,bar,][]{Al-Refaie+2021,Changeat_2022_fiveKey,Edwards+2023}
in standard atmospheric retrievals, but it is additionally coupled to the melt--atmosphere chemistry.
The assumption of an Earth-like iron mass fraction is motivated by statistical analyses of the observed exoplanet population \citep{Rogers+2021} and allows us to remove interior compositional degeneracies, while focusing on the coupling between the atmosphere and the magma ocean. 

In reality, however, the core radius can vary substantially depending on the amount of iron and volatiles stored in the planet.
We note that variations in the iron mass fraction can lead to planetary radius differences of a few tens of percent when comparing pure silicate planets to
Mercury-like iron core cases \citep[e.g.,][]{Seager+07,Fortney+2007}.
In addition, the incorporation of hydrogen and oxygen into the innermost iron core can further inflate the planetary radius, enhancing the rocky core radius by up to a few tens of percent even for an Earth-like
iron fraction \citep[see Fig.~16 of][]{Schlichting+2022}.
As a result, planets with identical masses may exhibit a non-negligible
range of possible core radii. Variations in the core radius would, in principle, lead to changes in the
atmospheric thickness, thereby affecting the inferred
temperature, pressure, and compositional constraints, including the mean molecular weight of the atmosphere.

In this work, we do not investigate these effects; however, future versions of \textsc{MELTYQ} could incorporate more detailed representations of the rocky core, including the retrieval of iron mass fraction and hydrogen inventory. In the present study, we emphasize that this simplifying assumption may introduce significant biases, and we therefore urge caution when interpreting the inferred atmospheric and interior constraints.

\begin{comment}
In the current implementation of MELTYQ, these
effects are absorbed into a fixed core radius for a given
mass, which may introduce systematic biases when interpreting atmospheric constraints in terms of interior
properties. A natural extension of this framework is
therefore to treat the core radius, or equivalently the
iron mass fraction and hydrogen inventory, as addi-
tional retrieval parameters. Incorporating these degrees
of freedom will enable a more self-consistent connection
between the retrieved atmospheric properties and the
planet’s interior structure, and will be explored in future
work. 
\end{comment}

\subsection{Equation of state limitations}
\label{sec:eos}
The quantitative accuracy of our framework is inherently limited by the ($T,P$) ranges over which the adopted EOS are valid.
In particular, the available EOS for NH$_3$ is quantitatively constrained
to a relatively narrow domain; for example, the current reference EOS for
 NH$_3$ is validated only up to $T = 725$~K and $P = 1$~GPa
\citep{Gao+2023}.
For this reason, we neglect non-ideal effects for nitrogen-bearing species,
and the resulting quantitative constraints involving these species are
therefore limited.
In addition, the EOS of \citet{Zhang+2009}, which we employ for H-, O-, and
C-bearing gases has been validated over a finite range of temperatures, $T \in [673,2573]$~K, and pressures, $P \in [0.001,10]$~GPa.
 We allow extrapolation toward higher temperatures and restrict
the pressure range in our framework.
While temperature extrapolation may introduce additional systematic
uncertainties, restricting the pressure range can also systematically
influence the explored parameter space by excluding solutions at higher
pressures.
Accordingly, constraints involving NH$_3$ and those inferred at higher
temperatures or near the upper pressure limits should be interpreted with
additional caution when the inferred conditions approach or exceed the
validated EOS domains.

In addition, EOS for volatile mixtures have been developed, but are typically validated over more limited ranges of temperature, pressure, and composition, which further constrains the quantitative accuracy of our framework that neglects mixture non-ideality.
Existing EOS for mixtures such as H$_2$-He, H$_2$O-CO$_2$, and H$_2$O-CH$_4$ have been explored at relatively high temperatures and pressures \citep{Chabrier+2021,Duan+2006,Zhang+2007}. However, EOS for mixtures of H$_2$ with C-, O-, or N-bearing species remain more limited, to our knowledge.
For example, EOS for binary hydrogen mixtures containing CH$_4$, N$_2$, CO, and CO$_2$ have been validated only up to $T = 700$~K and $P = 70$~MPa, with density uncertainties of 0.2--0.5~\% \citep{Kunz+2012,Beckmuller+2021}.

These EOS-related limitations are not specific to \textsc{MELTYQ}, but are
common to all magma--atmosphere coupling models developed for sub-Neptunes.
Future laboratory experiments and numerical studies aimed at improving equations of state and extending their range of temperatures,
pressures, and compositions 
 will therefore be
crucial for advancing magma--atmosphere coupling models, including those
applied to hydrogen-dominated atmospheres interacting with magma oceans
in sub-Neptunes.

\subsection{Temperature profile of deep atmospheric region}
\label{sec:tp}
The thermal structure of the deep atmospheric region is often expected to be governed primarily by an adiabatic lapse rate as a result of convection.
Although we adopt a simplified temperature profile in which the temperature varies linearly with $\log P$ between $T_\mathrm{melt}$ at $P_\mathrm{melt}$ and
$T_\mathrm{b}$ at $P_\mathrm{b}$, it is in principle possible to impose an adiabatic temperature profile in the deep atmosphere. 
We do not do this since convection in the deep atmosphere is not guaranteed.
As discussed in Sec.~\ref{sec:mix}, rocky vapors near the magma ocean surface can suppress convection 
\citep{Misener+2022,Misener+2023}.
In addition, even in the absence of significant rocky vapors, deep radiative layers may form depending on stellar irradiation, internal heat flux, and atmospheric opacity \citep{Thorngren+2019,Selsis+2023}.
Because observational constraints and theoretical models on the thermal structure at these depths remain limited, adopting a simplified temperature profile, as done in this study, represents a more practical approximation.
In the future, a more self-consistent treatment of the thermal structure, and/or an exploration of the impact of $T-P$ assumptions for the deep atmosphere could be done.

\subsection{Non-equilibrium chemistry in upper atmosphere}
\label{sec:eq}
Disequilibrium processes such as vertical mixing and photochemistry can play a dominant role in shaping the composition at upper atmospheric regions
\citep[e.g.,][]{Visscher+2011,Moses+2011,Venot+2012}.
In a forward melt-atmosphere coupling model reproducing the atmospheric composition constrained by the atmospheric retrieval of \citet{Benneke+2024} for TOI-270~d, \citet{Nixon+2025} considered  photochemistry and vertical mixing in the atmosphere. They found that 
the fractions of CO and CO$_2$ at $P\lesssim1$~bar depend on vertical mixing, while
the upper atmospheric fractions of H$_2$O and CH$_4$ are largely unaffected by vertical mixing and photochemistry. 
In particular, eddy diffusion coefficients larger than $10^4$ cm$^2$~s$^{-1}$ are required to reproduce the retrieved CO$_2$ abundance \citep[see Fig.~5 of][]{Nixon+2025}.
Such joint consideration of magma-atmosphere interactions and disequilibrium processes is therefore essential for quantifying the extent to which processes affect the abundances of individual atmospheric species.

In  the present work, we do not explicitly
account for these non-equilibrium processes and assumes a vertically uniform
composition in the upper atmospheric region above $P > 10$~bar.
This simplification is often done in standard atmospheric retrievals, and it is here motivated by our focus on demonstrating the
viability of our approach, but a more
comprehensive treatment of upper-atmospheric disequilibrium chemistry as done in retrievals from e.g., \cite{Kawashima_2021, Al-Refaie+2024, Bardet_2025} should be envisaged. Such an extension would enable a more realistic connection between the
observed atmospheric composition and the underlying interior properties,
as well as accounting for the impact from the atmospheric dynamics and the star.

In addition, sulfur-bearing species, which have been suggested to be present in the atmosphere of TOI-270~d from atmospheric retrieval works \citep[e.g.,][]{Holmberg_2024,Felix+2025}, are expected to be strongly influenced by photochemistry \citep{Tsai+2023,Veillet+2026}. While experimental constraints on sulfur solubility in silicate melts are available \citep[e.g.,][]{ONeill+2002,Boulliung2023}, a self-consistent treatment requires combining melt-atmosphere partitioning with atmospheric photochemistry. Since photochemistry is not included in the present model, sulfur chemistry is not considered in this work. Incorporating sulfur species in a physically consistent manner is therefore left for our future work.

\section{Summary}
\label{sec:summary}
In this article, we have introduced a novel coupled magma-atmosphere retrieval framework, \textsc{MELTYQ}, designed to connect atmospheric spectra of sub-Neptune exoplanets to the composition and redox state of underlying magma oceans. MELTYQ self-consistently links interior volatile inventories, gas-melt equilibrium, deep atmospheric structure, and observable transmission spectra within the Bayesian retrieval architecture of \textsc{TauREx}. This approach represents a conceptual and methodological advance over previous studies by embedding magma-atmosphere coupling directly into the retrieval process, enabling uncertainties and degeneracies to be quantitatively propagated from atmospheric data to interior properties.

After rigorously validating the concept using self-retrievals, we conducted tests on the JWST observations of K2-18\,b and TOI-270\,d. For K2-18\,b, we find that the observed spectrum can be well fit by a hydrogen-dominated atmosphere containing H$_2$O, CH$_4$, CO, and CO$_2$, 
in equilibrium with a deep magma ocean: the data place meaningful constraints on interior redox state and carbon content, albeit with strong degeneracies. For TOI-270\,d, we find evidence that the spectrum cannot be fully fitted by a \textsc{MELTYQ} retrieval, implying that disequilibrium chemistry or a breakdown of magma-atmosphere coupling is required to explain the observed CO/CO$_2$ feature. Overall, these tests indicate that magma composition parameters---including oxygen fugacity and volatile abundances---can, in principle, be statistically recovered from transmission spectra when the model assumptions are satisfied.

Importantly, our results emphasize that inferring {\it any} interior properties from atmospheric spectra relies on simplifying assumptions---which must always reflect our physical knowledge, but also importantly match the information content of the data---regarding interior structure, vertical mixing, thermal profiles, and equations of state. We have identified and discussed some of these assumptions, highlighting physical regimes where a \textsc{MELTYQ}-like approach might be reasonable, as well as those where it is not applicable, such as fully miscible interiors or strongly stratified atmospheres. Despite these limitations, this conceptual work demonstrates that embedding coupled atmosphere--interior physics in a Bayesian retrieval is possible, providing a powerful and transparent way to interpret sub-Neptune observations.

\begin{acknowledgments}
This work was supported by JSPS KAKENHI grant No. 25K01062. This publication is part of the project ``Exoplanet Atmospheres with Next-generation Space Telescopes'' with file no. VI.Veni.242.091 (PI: Q. Changeat) of the ``NWO Talent Programme Veni Science domain 2024'' under the grant \url{https://doi.org/10.61686/QPZSS86131}. It is also part of the project ``Interpreting exoplanet atmospheres with JWST'' with file no. 2024.034 (PI: Q. Changeat) of the research programme ``Rekentijd nationale computersystemen'' which is (partly) financed by the Dutch Research Council (NWO) under the grant \url{https://doi.org/10.61686/QXVQT85756}. 
\end{acknowledgments}

\section{Data Availability}
The JWST NIRISS Transmission Spectrum of K2-18\,b used in Madhusudhan et al. 2023 was obtained from the OSF \citep{madhu2023data}. The JWST NIRSpec Transmission Spectrum of K2-18\,b used in Hu et al. 2025 was obtained from OSF \citep{hu2025data}. The JWST MIRI Transmission Spectrum of K2-18\,b used in Madhusudhan et al. 2025 was obtained from OSF \citep{madhu2025data}. The JWST NIRSpec Transmission Spectrum of TOI-270\,d used in Holmberg\&Madhusudhan 2024 was obtained from OSF \citep{madhu2024data}.
Data products from this paper can be shared upon reasonable request to the corresponding author.
\bibliography{ref}{}
\bibliographystyle{aasjournal}

\appendix

\section{Solubility laws}\label{app:solub}
Many volatile species are soluble in magma. In this model, we consider the solubility effect of volatiles, especially H$_2$, H$_2$O, CO, CO$_2$, and N$_2$, which have been investigated in experimental studies. Below, we summarize the solubility laws we adopt.

For H$_2$O and CO$_2$, we adopt the solubility laws derived by \citet{lichtenberg+2021}, which is given by:
\begin{eqnarray}
\left(\frac{Y_\mathrm{H_2O}}{1.033\times10^{-6}} \right)^{1.747}=P_\mathrm{H_2O} [\mathrm{Pa}],
\label{eq:sol2}
\end{eqnarray}
\begin{eqnarray}
\left(\frac{Y_\mathrm{CO_2}} {1.937\times10^{-15}}\right)^{{0.714}}=P_\mathrm{CO_2} [\mathrm{Pa}],
\label{eq:sol3}
\end{eqnarray}
where ${Y_\mathrm{H_2O}}$ and ${Y_\mathrm{CO_2}}$ are the mass fractions of H$_2$O and CO$_2$ in melt, respectively.
For H$_2$, CH$_4$, and CO, we adopt the solubility laws used in \citet{Seo+2024}, which are based on the experimental works of \citet{Hirschmann+2012}, \citet{Ardia+2013}, and \citet{Yoshioka+2019}, respectively.
The formula of their solubility laws are given by
\begin{eqnarray}
X_{\rm H_2}=f_{H_{2}}[{\rm bar}]\ \exp({-11.403-0.76P_\mathrm{MELT}[{\rm GPa}]}),
\label{eq:sol1}
\end{eqnarray}
\begin{eqnarray}
X_{\rm CH_4}=f_{CH_{4}}[{\rm bar}]\ \exp({-7.63-1.9P_\mathrm{MELT}[{\rm GPa}]}),
\label{eq:solch4}
\end{eqnarray}
\begin{eqnarray}
Y_\mathrm{CO} = 10^{-7.2+0.8 \log_{10}(f_\mathrm{CO} [\mathrm{GPa}])},
\label{eq:solco}
\end{eqnarray}
where $X_{\rm H_2}$ and $X_{\rm CH_4}$ are the mole fractions of H$_2$ and CH$_4$ in melt, respectively, and $Y_\mathrm{CO}$ is the mass fraction of CO in melt. Note that the constant term in the exponent of Eq.~\ref{eq:solco} (i.e., $-7.2$) follows the value
reported by \citet{Yoshioka+2019}, which forms the basis of the formulation
used by \citet{Seo+2024}, and differs from the value explicitly written in
\citet{Seo+2024} by 10$^{0.37}$.
We adopt the former here, as the latter was confirmed to be an
error (C. Seo 2026, private communication).

Also, for N$_2$, we use the solubility laws derived by \citet{Dasgupta+2022}
\begin{eqnarray}
\left(\frac{Y_\mathrm{N_2}} {10^{-6}}\right) &=&
P_\mathrm{N_2}^{0.5}[\mathrm{GPa}]\exp\left( 5908 \frac{P_\mathrm{MELT}\mathrm{[GPa]}}{T}-1.6\Delta \mathrm{IW}\right) \nonumber \\
&& + P_\mathrm{N_2} [\mathrm{GPa}] \exp(4.67+7.11X_\mathrm{SiO_2} \nonumber \\
&& -13.06X_\mathrm{Al_2O_3} - 120.67X_\mathrm{TiO_2} ),
\label{eq:sol4}
\end{eqnarray}
where $Y_\mathrm{N_2}$ is the mass fraction of $N_2$ in melt, and $X_s$ is the molar fraction of rock species, $s$. Following \citep{Dasgupta+2022}, we assume basaltic melt compositions of SiO$_2$ (56\%), Al$_2$O$_3$ (11\%), and TiO$_2$ (1\%). Although the fractions of volatile species are expressed either as mass fractions or as molar fractions depending on the formulation described above, we use all volatile abundances as molar fractions for clarity and consistency. To this end, we assume a representative mean molar mass of 60 g/mol for silicate melts, whereas the value derived from the mean composition of mid-ocean ridge basalts is 62 g/mol \citep{Gale+2013}.

\newpage
\section{Numerical test of the magma--atmosphere coupling code}
\label{app:bench}

We perform a numerical test of the magma--atmosphere coupling code
developed in this study.
As described below, we confirm that our code successfully reproduce
the volatile speciation between the silicate melt and the atmosphere for a
sub-Neptune shown in \citet{Seo+2024}.

\citet{Seo+2024} calculated the speciation of H-, O-, and C-bearing species
between an atmosphere and a magma ocean for prescribed values of
$T_\mathrm{MELT}$, $P_\mathrm{MELT}$, the composition of the accreted gas, and the number ratio between the oxygen bound to iron and iron itself before reaction  ($N_\mathrm{O(Fe)}^\mathrm{before}/N_\mathrm{Fe}$), accounting for both volatile
dissolution and redox reactions.
Although some of their input parameters differ from those adopted in this work (i.e., $N_\mathrm{MELT}$, $\mathrm{CO}_\mathrm{MELT}$ and 
$f_{\mathrm{O}_2}$), we adopt their
resulting values of $f_{\mathrm{O}_2}$ and $\mathrm{CO}_\mathrm{MELT}$ and set
$N_\mathrm{MELT}=0$ for the numerical benchmark test presented here. In addition, for this numerical test only, we adopt Eqs.~9, 13, and 15 of
\citet{Seo+2024} for the solubility laws of H$_2$O, CO$_2$, and CO,
respectively, instead of Eqs.~(\ref{eq:sol2}), (\ref{eq:sol3}), and
(\ref{eq:sol4}), in order to ensure full consistency with \citet{Seo+2024}.

\begin{figure*}[b]
  \begin{minipage}{0.5\textwidth}
    \begin{center}   
    ($a$) Volatile fraction in atmosphere
\includegraphics[width=\textwidth,trim=0cm 0cm 0cm 0.5cm,clip]{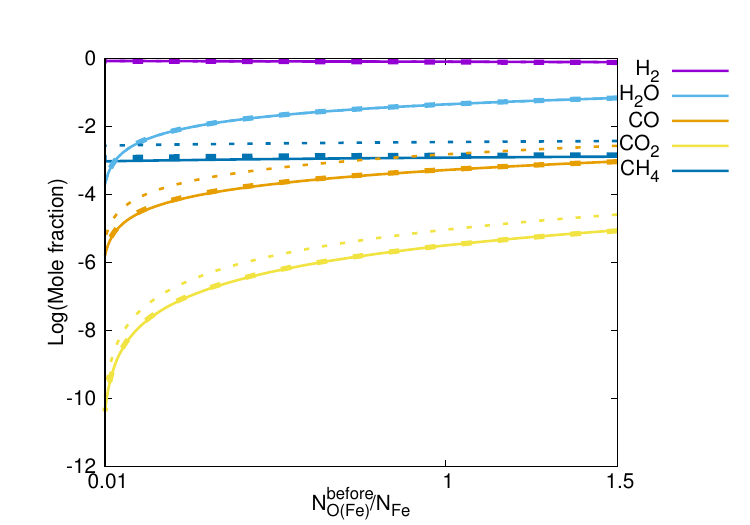}
   \end{center}
 \end{minipage}
   \begin{minipage}{0.5\textwidth}
    \begin{center}   
   ($b$) Volatile fraction in magma ocean
\includegraphics[width=\textwidth,trim=0cm 0cm 0cm 0.5cm,clip]{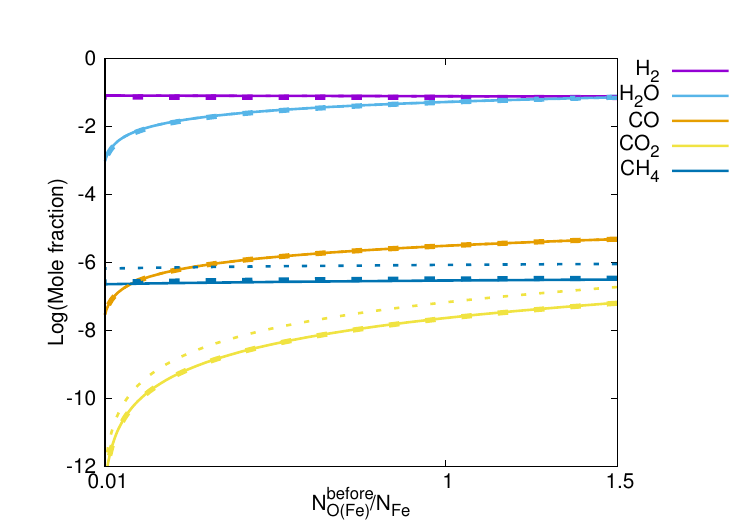}
   \end{center}
 \end{minipage}
  \caption{Reproduction of the speciation of H-, O-, and C-bearing species between an atmosphere and a magma ocean shown in \citet[dashed lines]{Seo+2024} that assume $T_\mathrm{MELT}=3000$~K and $P_\mathrm{MELT}=1$~GPa. The fraction of H$_2$ (purple), H$_2$O (cyan), CO (orange), CO$_2$ (yeallow), and CH$_4$ (blue) in an atmosphere ($a$) and a magma ocean ($b$) are shown as functions of $N_\mathrm{O(Fe)}^\mathrm{before}/N_\mathrm{Fe}$ (see text for the detail). Solid lines show our calculations using the same solubility laws adopted by
\citet{Seo+2024}, while dotted lines show our calculations using the CO
solubility law given by Eq.~\ref{eq:solco}.
}
\label{fig:atom_rep}
\end{figure*}

In Fig.~\ref{fig:atom_rep}, we reproduce the results of \citet{Seo+2024}.
Figure~\ref{fig:atom_rep} shows the fractions of H-, O-, and C-bearing species
in the atmosphere (Fig~\ref{fig:atom_rep}$a$) and in the magma ocean (Fig~\ref{fig:atom_rep}$b$) for
$T_\mathrm{MELT}=3000$~K and $P_\mathrm{MELT}=1$~GPa.
These results can be directly compared with those of the previous study
(dashed lines), shown in Fig.~4 of \citet{Seo+2024}.
Overall, our results are consistent with those of \citet{Seo+2024}, although
differences of up to $\sim20\%$ are found, particularly in the atmospheric
CH$_4$ fraction.
These differences likely arise from differences in the treatment of thermodynamic data and the computation of equilibrium constants. In this work, we derive Gibbs free energies and equilibrium constants using the JANAF database, whereas \citet{Seo+2024} actually adopted analytic expressions based on \citet{Posada+2005}, \citet{Ortenzi+2020}, and \citet{Kite+2020}, which leads to quantitative discrepancies (C. Seo 2026, private communication).

As described in Appendix~\ref{app:solub}, we have pointed out an error in the CO
solubility law shown in \citet{Seo+2024}.
Accordingly, we perform our simulations using the corrected CO solubility law
given by Eq.~\ref{eq:solco}.
The resulting calculations are shown as dotted lines in
Fig.~\ref{fig:atom_rep}, demonstrating that the error leads to an
approximately factor-of-two difference in the abundances of C-bearing
species in this setup.
This is consistent with the difference implied by the two formulations of the CO solubility law,
which differ by a factor of $10^{0.37}(\sim2.3)$. 

We note that the main findings of \citet{Seo+2024} would still hold even if the CO solubility law given by Eq.~\ref{eq:solco} were adopted.
In particular, the negative O/H--C/O trend and the depletion of C/O ratios under oxidized conditions, which arise from the solubility difference between
H- and C-bearing species, remain valid.
This is because \citet{Seo+2024} has overestimated the solubility of CO in the magma ocean relative to Eq.~\ref{eq:solco}, thereby underestimating the solubility difference between H- and C-bearing species.

\newpage

\section{Parameter dependence of our magma--atmosphere coupling model} \label{app:para}

\begin{figure*}
  \begin{minipage}{0.5\textwidth}
    \begin{center}   
    ($a$) Dependence on $P_\mathrm{MELT}$
\includegraphics[width=\textwidth,trim=0cm 0cm 0cm 0.5cm,clip]{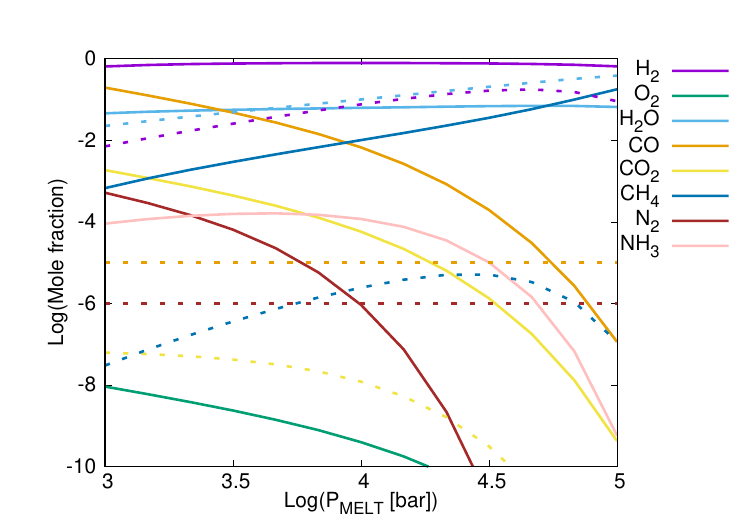}
   \end{center}
 \end{minipage}
   \begin{minipage}{0.5\textwidth}
    \begin{center}   
   ($b$) Dependence on $T_\mathrm{MELT}$
\includegraphics[width=\textwidth,trim=0cm 0cm 0cm 0.5cm,clip]{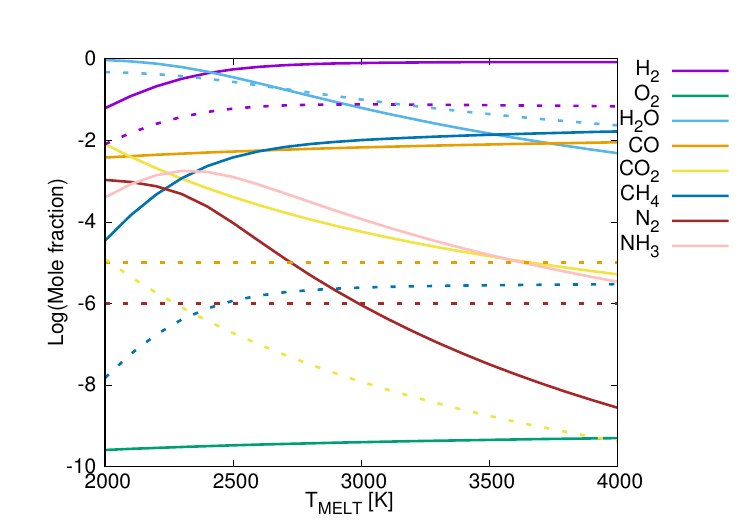}
   \end{center}
 \end{minipage}
   \begin{minipage}{0.5\textwidth}
    \begin{center}   
   ($c$) Dependence on $f_\mathrm{O_2}$
\includegraphics[width=\textwidth,trim=0cm 0cm 0cm 0.5cm,clip]{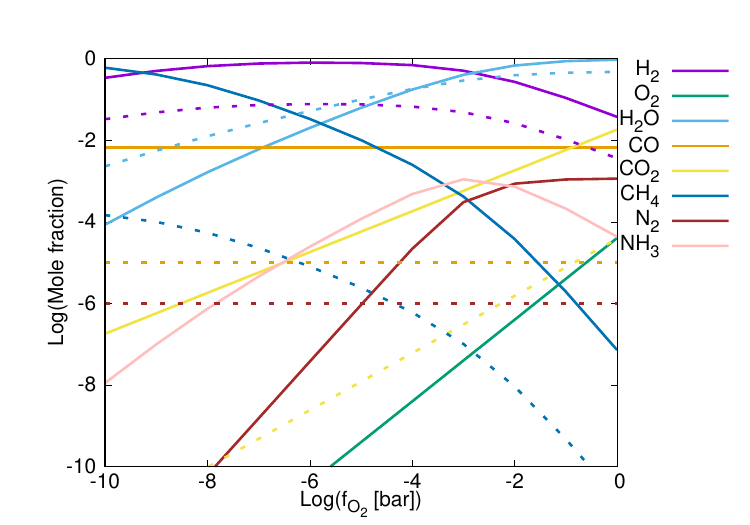}
   \end{center}
 \end{minipage}
   \begin{minipage}{0.5\textwidth}
    \begin{center}   
  ($d$) Dependence on CO fraction in melt
\includegraphics[width=\textwidth,trim=0cm 0cm 0cm 0.5cm,clip]{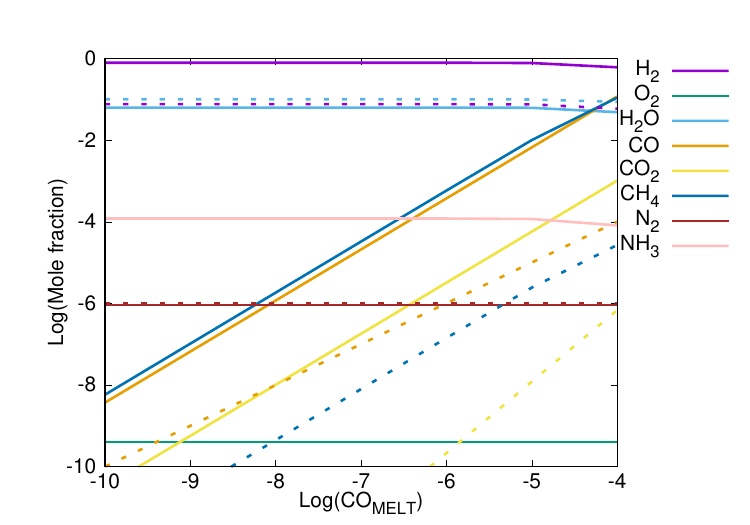}
   \end{center}
 \end{minipage}
  \begin{minipage}{0.5\textwidth}
    \begin{center}   
  ($e$)  Dependence on N$_2$ fraction in melt
\includegraphics[width=\textwidth,trim=0cm 0cm 0cm 0.5cm,clip]{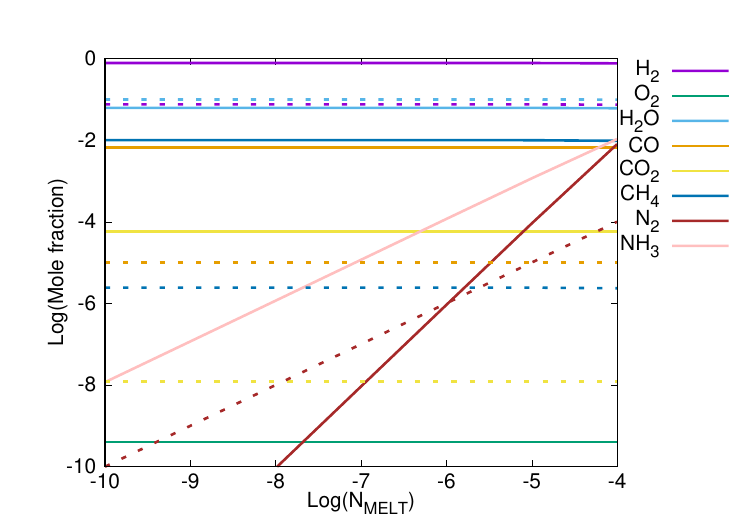}
   \end{center}
 \end{minipage}
 \begin{minipage}{0.5\textwidth}
    \begin{center}   
   ($f$) Dependence on $T_\mathrm{b}$
\includegraphics[width=\textwidth,trim=0cm 0cm 0cm 0.5cm,clip]{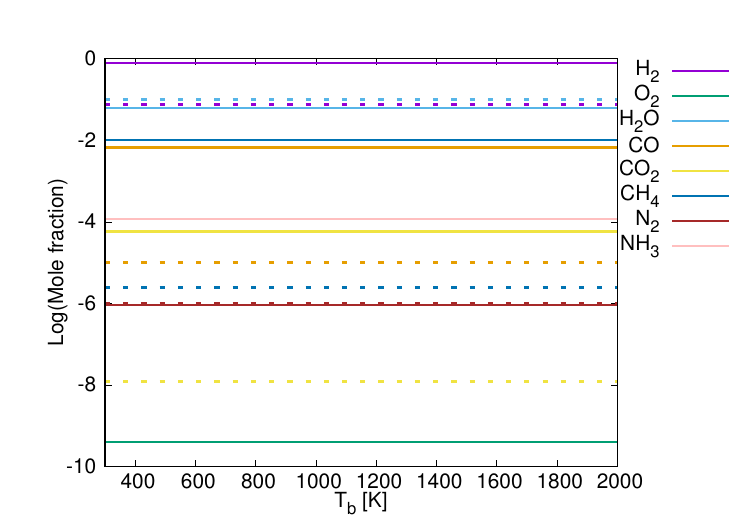}
   \end{center}
 \end{minipage}

\caption{
Molar fractions of gas species at $P_\mathrm{melt}$ (solid lines) and those in melt (dotted lines) are shown 
as functions of (a) $P_\mathrm{melt}$, (b) $T_\mathrm{melt}$, (c) $f_{\mathrm{O}_2}$,
(d) CO abundance in the melt, (e) N$_2$ abundance in the melt, and (f) $T_\mathrm{b}$.
Unless varied along the x-axis, the parameters are fixed at:
planet mass of $4\,M_\oplus$, $P_\mathrm{melt}=10^4$~bar, $T_\mathrm{melt}=3000$~K,
$f_{\mathrm{O}_2}=10^{-5}$~bar, CO abundance in the melt of $10^{-5}$,  
N$_2$ abundance in the melt of $10^{-6}$, and $T_\mathrm{b}=1000$~K.
Colors indicate individual species:  
H$_2$ (purple), O$_2$ (green), H$_2$O (cyan), CO (orange), CO$_2$ (yeallow), CH$_4$ (blue),  
N$_2$ (brown), and NH$_3$ (pink). Note that He is not shown in the plots, but it is included in the calculations by assuming an H--He mixture with a helium mass fraction of 0.275.
}
\label{fig:atom_compx}
\end{figure*}

\begin{figure*}
  \begin{minipage}{0.5\textwidth}
    \begin{center}   
    ($a$) Dependence on $P_\mathrm{MELT}$
\includegraphics[width=\textwidth,trim=0cm 0cm 0cm 0.5cm,clip]{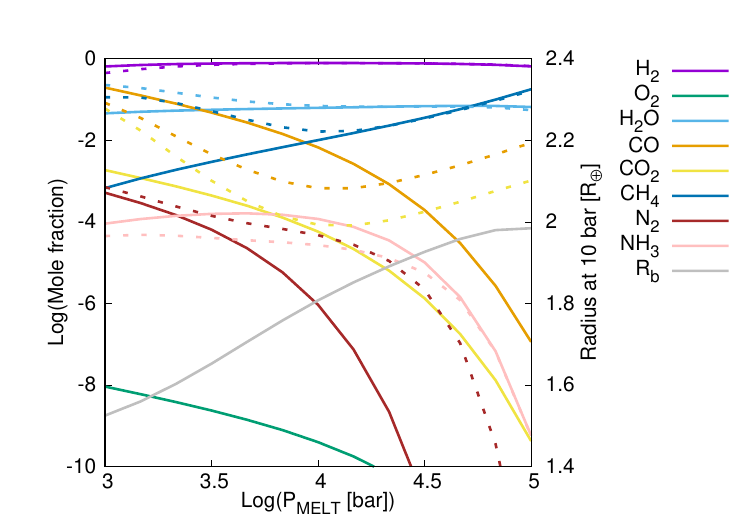}
   \end{center}
 \end{minipage}
   \begin{minipage}{0.5\textwidth}
    \begin{center}   
   ($b$) Dependence on $T_\mathrm{MELT}$
\includegraphics[width=\textwidth,trim=0cm 0cm 0cm 0.5cm,clip]{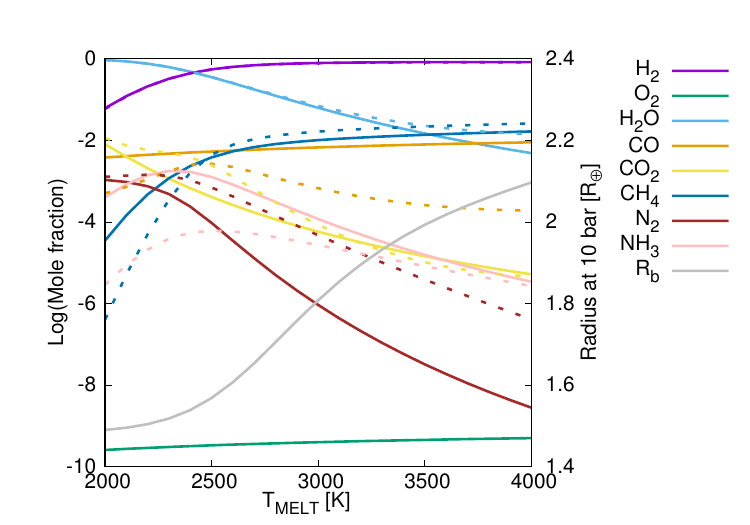}
   \end{center}
 \end{minipage}
   \begin{minipage}{0.5\textwidth}
    \begin{center}   
   ($c$) Dependence on $f_\mathrm{O_2}$
\includegraphics[width=\textwidth,trim=0cm 0cm 0cm 0.5cm,clip]{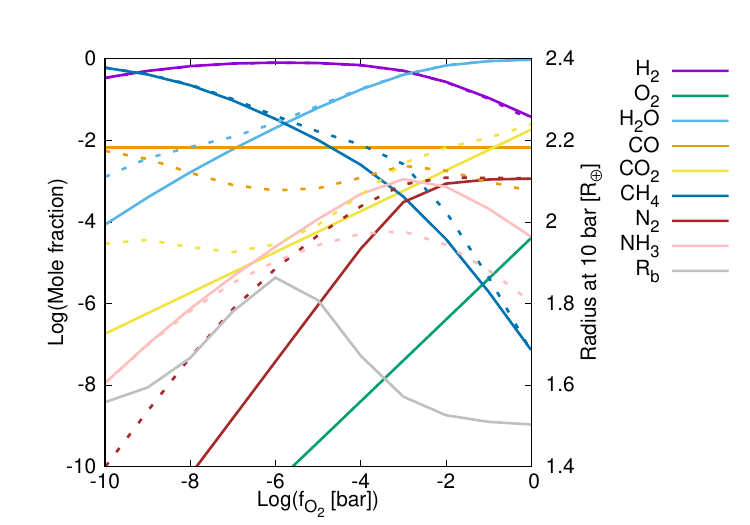}
   \end{center}
 \end{minipage}
   \begin{minipage}{0.5\textwidth}
    \begin{center}   
  ($d$) Dependence on CO fraction in melt
\includegraphics[width=\textwidth,trim=0cm 0cm 0cm 0.5cm,clip]{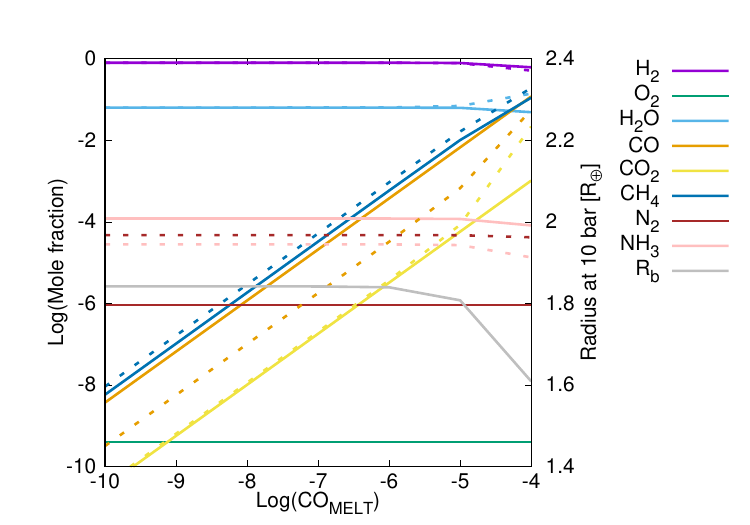}
   \end{center}
 \end{minipage}
  \begin{minipage}{0.5\textwidth}
    \begin{center}   
  ($e$)  Dependence on N$_2$ fraction in melt
\includegraphics[width=\textwidth,trim=0cm 0cm 0cm 0.5cm,clip]{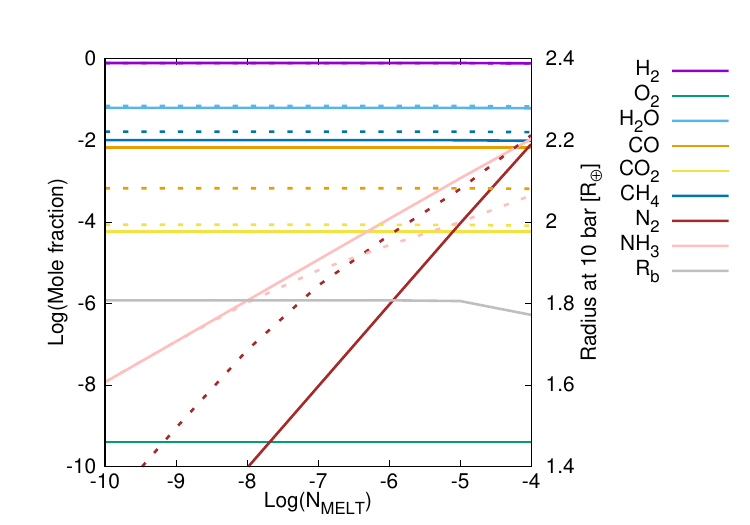}
   \end{center}
 \end{minipage}
 \begin{minipage}{0.5\textwidth}
    \begin{center}   
   ($f$) Dependence on $T_\mathrm{b}$
\includegraphics[width=\textwidth,trim=0cm 0cm 0cm 0.5cm,clip]{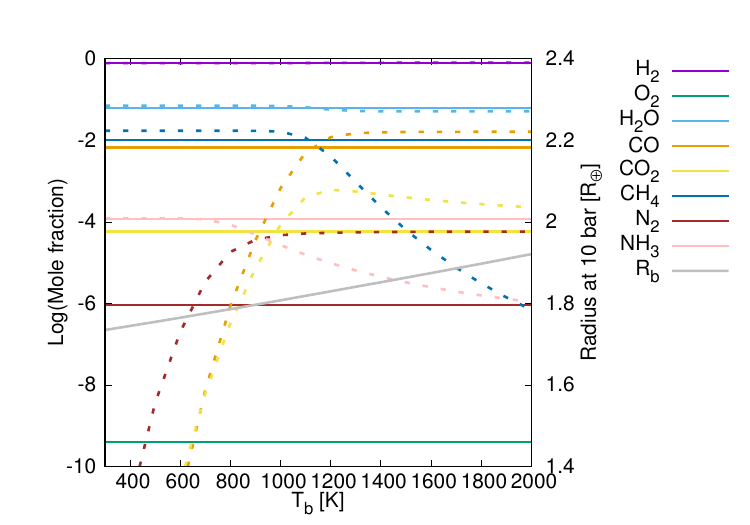}
   \end{center}
 \end{minipage}

\caption{
Molar fractions of gas species at $P_\mathrm{melt}$ (solid lines) and $P_\mathrm{b}$ (dotted lines), together with the radius at 10~bar (solid gray line), are shown
as functions of (a) $P_\mathrm{melt}$, (b) $T_\mathrm{melt}$, (c) $f_{\mathrm{O}_2}$,
(d) CO abundance in the melt, (e) N$_2$ abundance in the melt, and (f) $T_\mathrm{b}$.
The parameter setting is as same as Fig.~\ref{fig:atom_compx}.
}
\label{fig:atom_compy}
\end{figure*}

Here, we show the parameter dependence of our magma--atmosphere coupling
model in Figures~\ref{fig:atom_compx} and \ref{fig:atom_compy}.
As described in Sec.~\ref{sec:numpro}, for the magma ocean and deep atmospheric
regions, our model calculates the fractions of H$_2$, H$_2$O, CO$_2$, and
CH$_4$ in the silicate melt, the atmospheric composition between
$P_\mathrm{MELT}$ and $P_\mathrm{b}$ (corresponding to 10~bar in this study) from six input
parameters: $P_\mathrm{MELT}$, $T_\mathrm{MELT}$, $f_{\mathrm{O}_2}$,
$\mathrm{CO}_\mathrm{MELT}$, $N_\mathrm{MELT}$, and $T_\mathrm{b}$.

Figure~\ref{fig:atom_compx} shows the atmospheric composition at
$P_\mathrm{MELT}$ (solid lines) together with the volatile fractions in the
melt (dotted lines).
Figure~\ref{fig:atom_compy} shows the atmospheric composition at
$P_\mathrm{MELT}$ (solid lines) and at $P_\mathrm{b}$ (dotted lines) together with the planetary radius at
$P_\mathrm{b}$, $R_\mathrm{b}$ (gray dotted lines).
In both figures, each panel illustrates the dependence on one of the input
parameters: $P_\mathrm{MELT}$ ($a$), $T_\mathrm{MELT}$ ($b$),
$f_{\mathrm{O}_2}$ ($c$), $\mathrm{CO}_\mathrm{MELT}$ ($d$),
$N_\mathrm{MELT}$ ($e$), and $T_\mathrm{b}$ ($f$).
Unless varied along the horizontal axis, the parameters are fixed at a
planetary mass of $4\,M_\oplus$, $P_\mathrm{MELT}=10^4$~bar,
$T_\mathrm{MELT}=3000$~K, $f_{\mathrm{O}_2}=10^{-5}$~bar,
$\mathrm{CO}_\mathrm{MELT}=10^{-5}$, $N_\mathrm{MELT}=10^{-6}$,
and $T_\mathrm{b}=1000$~K.

\section{Posterior distribution of MELTYQ for K2-18\,b} \label{app:post}

\begin{figure*}
\centering
    \includegraphics[width = 0.98\textwidth]{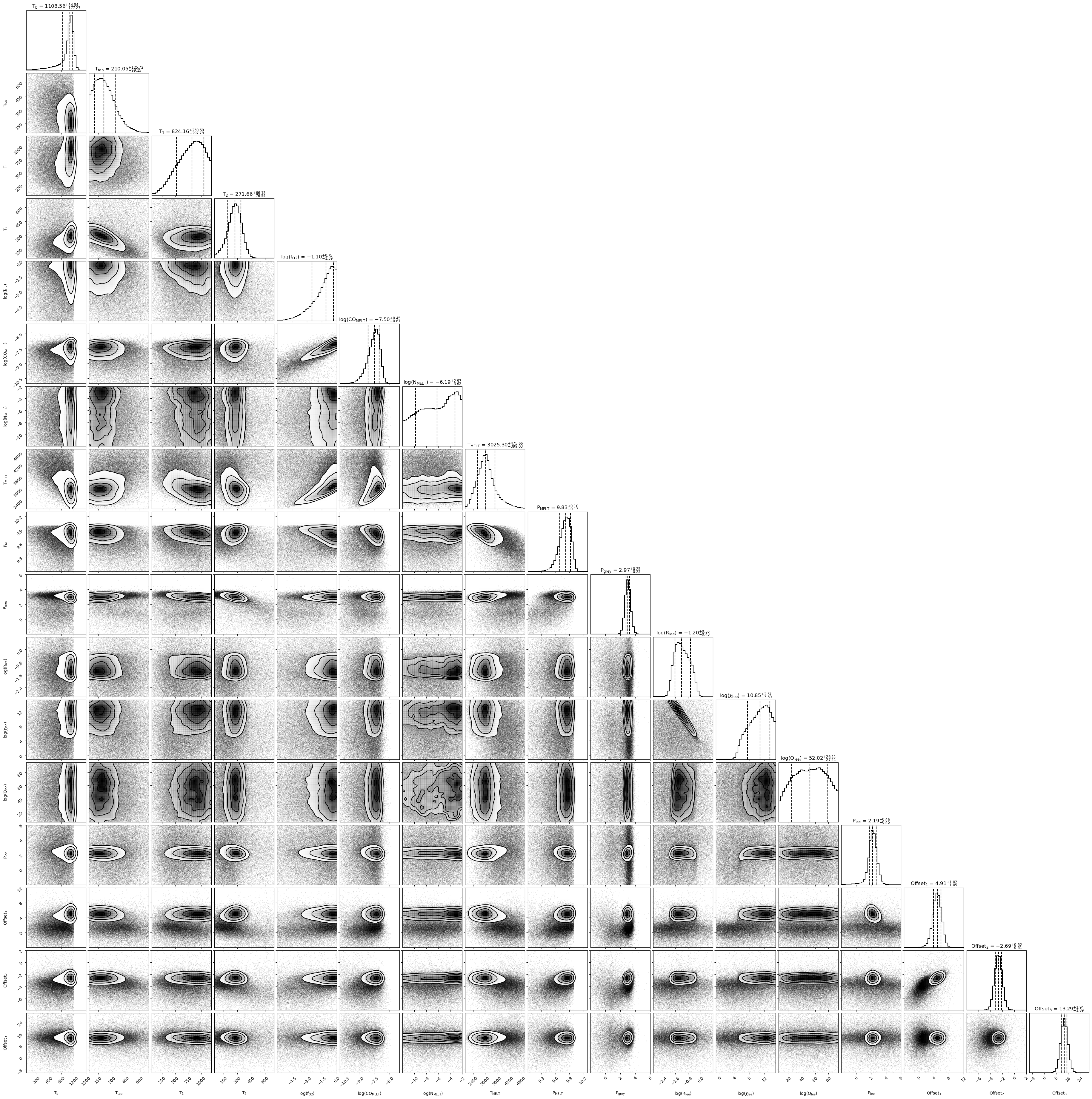}
    \caption{Full posterior distribution of the MELTYQ retrieval for the K2-18\,b data.}\label{fig:corner_k218}
\end{figure*}

Figure \ref{fig:corner_k218} shows the posterior distributions of the \textsc{MELTYQ retrieval for K2-18\,b}.

\end{document}